\documentclass[fleqn,usenatbib]{mnras}

\usepackage{newtxtext,newtxmath}
\usepackage{natbib}
\usepackage{stfloats}
\usepackage{soul}
\usepackage{lipsum}  
\usepackage{multirow}

\usepackage[T1]{fontenc}

\DeclareRobustCommand{\VAN}[3]{#2}
\let\VANthebibliography\thebibliography
\def\thebibliography{\DeclareRobustCommand{\VAN}[3]{##3}\VANthebibliography}

\usepackage{graphicx}	
\usepackage{amsmath}	
\usepackage[list=no]{subcaption}
\usepackage{float}
\usepackage{xcolor}
\usepackage{comment} 

\usepackage{amsmath}

\title[Barred LSB galaxies in TNG100]{ Study of barred galaxies in IllustrisTNG100: the case of low surface brightness galaxies}

\author[Chim Ramirez et al.]{
Karol Chim-Ramirez,$^{1}$\thanks{E-mail: k.chim@irya.unam.mx}
Bernardo Cervantes-Sodi,$^{1}$
Yetli Rosas-Guevara,$^{2}$
Luis Enrique Pérez-Montaño$^{3,1}$ 
\newauthor  and Silvia Bonoli$^{2,4}$
\\
$^{1}$Instituto de Radioastronomía y Astrofísica, Universidad Nacional Autónoma de México, Antigua Carretera a Pátzcuaro 8701, Ex-Hda. San José \\ de la Huerta, Morelia, Michoacán, México C.P. 58089\\
$^{2}$Donostia International Physics Centre (DIPC), Paseo Manuel de Lardizabal 4, 20018 Donostia-San Sebastian, Spain  \\
$^{3}$Institute for Astronomy, School of Physics, Zhejiang University, Hangzhou 310027, People's Republic of China  \\
$^{4}$IKERBASQUE, Basque Foundation for Science, E-48013, Bilbao, Spain\\
}

\date{Accepted XXX. Received YYY; in original form ZZZ}

\pubyear{2025}

\begin{document}
\label{firstpage}
\pagerange{\pageref{firstpage}--\pageref{lastpage}}
\maketitle

\begin{abstract}
In this work, we compare the presence of stellar bars in low and high surface brightness galaxies (LSBs and HSBs, respectively) using the TNG100 simulation of the IllustrisTNG project. The sample consists of 4,244 disc galaxies at $z=0$ with stellar mass M$\star \geq 10^{10}$ M$\odot$. We find a bar fraction of $24 \pm 1.73 \%$ in LSBs, similar to the $28 \pm 0.74\%$ found in HSBs, consistent with observations. For a given stellar mass range, HSBs consistently exhibit a higher bar fraction compared to LSBs, except at M$\star > 10^{11}$ M$\odot$, where the difference vanishes. To explore the origin of this trend and its relation to host galaxy properties, we construct several control samples matched in stellar mass, spin, gas mass fraction, and bulge-to-total mass ratio. For galaxies with M$\star<10^{11}$ M$\odot$, the lower bar fraction in LSBs appears to be associated with their higher spin and gas content—factors known to inhibit bar formation and growth. At the high mass end, only the bulge-to-total mass ratio is capable of enhancing the bar fraction difference, although its effect is limited. We also study the role of the local environment through the tidal parameter. Our results suggest that, unlike in HSBs, where the bar fraction remains largely unaffected, tidal interactions may promote bar formation in LSBs, albeit with a smaller impact than the intrinsic physical properties. These findings provide insight into the physical conditions that shape the presence of bars in LSBs.

\end{abstract}




\section{Introduction}

Stellar bars are non-axisymmetric components located in the inner part of disc galaxies. These structures play a significant role in the secular evolution of galaxies (\citealt{kormendy2013secular, debattista2004bulges}) as they can redistribute mass and angular momentum among different galaxy components, including the dark matter halo, the bulge, and the stellar disc (e.g. \citealt{lynden1979mechanism, sellwood1980galaxy, athanassoula2003determines}). Bars are found preferentially in massive galaxies with low-velocity dispersion, suggesting that they are dominated by rotation (\citealt{sheth2012hot}). These conditions promote the growth of the bar, pushing gas and stars into the central region of the galaxy, resulting in pseudobulge formation (\citealt{athanassoula2005nature}) and star formation in the galactocentric region (\citealt{khoperskov2018bar}).

Approximately two-thirds of nearby large spiral galaxies possess a bar, while the remaining third do not (e.g. \citealt{gavazzi2015role}).  Nonetheless, one unresolved question pertains to the conditions necessary for a disc galaxy to possess and maintain a bar structure. One method to investigate this observationally is using bar fractions, defined as the ratio of the number of bars to the number of discs as a function of certain galaxy properties. One primary property related to the presence of a bar in disc galaxies is the stellar mass. Numerous observational investigations have shown the correlation between bar fraction and stellar mass \citep{masters2012galaxy,diaz2016characterization, sodi2017low}. For instance, \cite{nair2010fraction}, using the catalogue presented in \cite{nair2010catalog} of $\sim 14,000$ galaxies from SDSS Data Release 4, have found that the bar fraction is strongly correlated with stellar mass.

Besides stellar mass, other galaxy properties including gas content, dark matter halo, angular momentum retention and bulge properties, have an important influence on the bar fraction \cite[e.g][]{weinzirl2009bulge, lee2012dependence, sodi2017stellar, kruk2018galaxy, romeo2023specific}. The gas content within a galaxy significantly influences the formation and evolution of a stellar bar. Specifically, \cite{masters2012galaxy} and \cite{sodi2017stellar} analyzed $2,090$ disc galaxies (from Galaxy Zoo 2 and ALFALFA) and $1, 471$  late-type galaxies from Sloan Digital Sky Survey Data Release 7 (SDSS-DR7), respectively. They found that the bar fraction in a sample is significantly lower in galaxies with high atomic hydrogen gas content, even at fixed stellar mass. Furthermore, \cite{sodi2017stellar} have found that the strong bar fraction decreases with increasing HI content, although a slight rise in the bar fraction of weak bars was noted. The properties of the bulge also appear to influence the presence of a bar. \cite{weinzirl2009bulge} determine the bar fraction using H-band images and find that barred galaxies have bulges with bulge-to-total mass fractions $B/T<0.2$ and smaller Sersic indexes ($n<2$). \cite{kruk2018galaxy} report analogous findings using the Galaxy Zoo survey.  \cite{cervantes2013galactic}, analyzing $\sim 10, 000$ disc galaxies from the SDSS, proposed an order-of-magnitude proxy for the galactic spin parameter, $\lambda_{\rm d}$, based on structural parameters of the stellar discs and a ratio of stellar-to-halo mass derived from the stellar surface density. The authors found that the galactic spin distribution of unbarred galaxies was statistically distinct. Galaxies with long bars present the lowest $\lambda_{\rm d}$ values, followed by unbarred galaxies, whereas galaxies with short bars display the highest $\lambda_{\rm d}$ values.

Theoretical models, in particular N-body simulations, have provided insights into the relation between galaxy properties and the presence of a bar.  \cite{laurikainen2004bar, jogee2005central} found that in galaxies with significant gas content, the angular momentum redistribution causes the gas to fall towards the central region, triggering a starburst, thus enhancing the central mass concentration which subsequently weakens the bar. This will result in the accelerated processing and depletion of gas in barred galaxies, potentially elucidating the prevalence of bars in gas-deficient galaxies. \cite{athanassoula2013bar, combes2011pattern} also show that bars in gas-rich galaxy models grow slowly and, at the end of their growth, the bar is weaker than in poor-gas galaxy models. In addition,  \cite{sellwood2001stability} have shown that dense centers could stabilize the disc against the modes characteristic of bar formation (\citealt{toomre1981amplifies}). A possible explanation is given by \cite{kataria2018study} who found through N-body simulations, that massive bulges increase the velocity dispersion in the disc and avoid the angular momentum exchange between the disc and halo making the disc stable.  Furthermore, they identified an upper limit of the mass fraction between the bulge and disc from which bars can no longer be formed, being lower for models with denser bulges. Consequently, bars are found preferentially in galaxies with less concentrated bulges. Furthermore, N-body simulation studies such as \cite{long2014secular}  suggest that lower spin favors the bar growth, whereas a spin parameter $\lambda> 0.03$ results in no growth. The authors also found that bar size has an anti-correlation with $\lambda$, attributable to the bar incapacity to catch orbits and growth in strength and size throughout the secular period.

Cosmological hydrodynamical simulations have enabled the investigation of bar structures within a cosmological framework. \cite{scannapieco2012bars} employ zoom-in hydrodynamical simulations to study bar properties in two Milky-Way-like galaxies, revealing that a massive bulge correlates with a larger and stronger stellar bar is larger and stronger compared to galaxies lacking a significant bulge component. This may result from the negligible angular momentum exchange in the model without the predominant bulge. In contrast, \cite{bonoli2016black} using \textit{ ERIS} (\citealt{guedes2011forming}), a zoom-in hydrodynamical simulation of a Milky Way halo,  find that the incorporation of AGN feedback reduces the size and mass of the bulge, favouring the formation of a bar. Certain studies suggest that interactions including mergers or flybys can delay the time of bar formation (if the structure is present), although, once the bar is formed, interactions do not affect the subsequent evolution of its properties (\citealt{zana2018external}). Recently, \cite{fragkoudi2020chemodynamics, fragkoudi2024bar}, using the Auriga simulations (\citealt{grand2017auriga}), highlight the varied properties of barred galaxies, both for the Milky Way and external barred galaxies, as well as the tight interplay between the central regions of galaxies (including the bulge), the bar, and the disc.

The investigation into the evolutionary patterns and characteristics of barred galaxies within a statistical framework has been made feasible by the advent of advanced cosmological hydrodynamic simulations (\citealt{vogelsberger2014introducing}, \citealt{schaye2015eagle}, \citealt{pillepich2018simulating}, \citealt{nelson2018first} review in \citealt{vogelsberger2020cosmological}, \citealt{crain2023hydrodynamical}). \cite{algorry2017barred} examined the EAGLE simulation and noted a notable decrease in the velocity of bars as they underwent evolutionary changes. This deceleration led to the enlargement of the inner areas of the dark matter halo. The findings of \cite{rosas2020buildup} and \cite{reddish2022newhorizon} indicate that barred galaxies have lower star formation rates and a greater gas deficiency than unbarred galaxies in the Illustris TNG100 and NewHorizon cosmological simulations (see also \citealt{zhao2020barred, zhou2020barred, rosas2022evolution, izquierdo2022disc}). Before the formation of the bar component, barred galaxies assembled their disc and bulge components at an earlier stage compared to unbarred galaxies, according to a study of their temporal evolution \cite{rosas2020buildup,izquierdo2022disc}, also showing that once the bar is fully settled, the gas content decreases, especially in the central part, producing a gas deplated galaxy nucleous and a higher percentage of quenched galaxies compared to unbarred galaxies (\citealt{lu2024illustristng}). Finally, employing the Illustris-1 simulation, \cite{peschken2019tidally} demonstrates that the preponderance of bars is triggered by external perturbations, such as mergers or flybys (also see \citealt{rosas2024rise} for TNG50). As a result of tidal forces, these interactions have the potential to cause the disintegration of bars. The authors also note that the presence of gas within the disc could impede the formation of tidally generated bars and reduce or eliminate existing bars. Although works such as \cite{bi2022emergence, bi2022modeling}, which use a series of high-resolution zoom-in simulations of galaxies with $z \sim 9-2$, have shown that the situation could be different at high redshifts, where galaxies are gas-rich and interactions could play a crucial role in bar formation. According to these studies, bars that form at high redshifts result from interactions such as mergers and flybys and are gas-rich, a result that could impact the bar fraction at low redshift, if correctly accounted for.

Given that low surface brightness galaxies (LSBs) are characterised by elevated spin parameters and substantial gas-to-stellar mass ratios, it is expected that they should present a low likelihood of hosting bars. Additionally, they are situated in sparsely populated environments (\citealt{mcgaugh2001high, sodi2017low, perez2019environment}). LSBs constitute a significant portion of the galaxy population ($\sim$ 50 \% according to \citealt{mcgaugh1995galaxy}), making them the biggest reservoirs of baryonic matter in the Universe (\citealt{impey1996low, o2000red}). These galaxies are often classified as disc galaxies with a central surface brightness in the B-band greater than $\mu_0 (B) \geq 22.7$ magarcsec$^{-2}$ (\citealt{freeman1970disks}). In this work, we will adopt the definition provided by \cite{di2019nihao}, who defines a low surface brightness (LSB) galaxy as one with a surface brightness value in the r-band $\mu_0 (r) \geq 22$ mag/arcsec$^2$. The use of this band is considered since it allows us to sample the underlying stellar mass of galaxies, instead of the most recent star formation traced by bluer bands.

LSBs are primarily distinguished by their low stellar surface mass density, in contrast to their high surface brightness (HSB) counterparts.  Their rotation curves slowly rise as they are dominated by dark matter (DM) at all radii (\citealt{de2001low, perez2019environment, zhu2023giant, kumar2023study}). These galaxies are formed in the center of dark matter halos characterised by high angular momentum (\citealt{dalcanton1997formation, mo1998formation}) with spin values above $\lambda > 0.05$ (\citealt{boissier2003chemical}). LSBs are expected to show more stable discs because of their sparse stellar discs encircled by rapidly rotating dark matter halos (\citealt{mihos1997dynamical, yurin2015stability}).  The elevated gas level in the LSBs would inhibit the formation and evolution of bars. The aforementioned properties of these galaxies indicate that these systems are suboptimal for bar formation. Indeed, \cite{sodi2017low} have calculated the surface brightness in the $B$-band, finding a bar fraction of $20 \%$ and $30 \%$ for LSBs and HSBs, respectively. In addition, the study by \cite{honey2016near} analyzed infrared images of LSBs obtained from SDSS, revealing that merely $\sim 8 \%$ of LSBs exhibited stellar bars, which may be ascribed to alternative mechanisms of bar formation, such as interactions or mergers. In this work, we will focus on the presence of a stellar bar in these two types of galaxies to gain some insights into the presence of bars in disc galaxies.

Our purpose is to examine the influence of galaxy properties on the existence of a bar through cosmological hydrodynamic simulations. Specifically, we study the bar fraction in LSBs and HSBs as a function of different galaxy properties (gas content, stellar mass, spin parameter, and bulge to total mass fraction) and environment. We additionally examine subsamples of LSBs and HSBs with analogous characteristics to assess their impact on the bar fraction.  We concentrate solely on the bar fraction to replicate observational studies that regard the bar fraction as an essential attribute for comprehending the evolution of bars.  To achieve this objective, we utilise the cosmological simulation TNG100 at redshift z=0.  The process for constructing our galaxy sample is detailed in Section \ref{methodology}, our primary results are outlined in Section \ref{results}, a general discussion of our results in the context of previous works is presented in Section \ref{discussion} and the final conclusions are provided in Section \ref{conclusions}.

\section{Methodology}
\label{methodology}
\subsection{IllustrisTNG Simulation}

The Illustris The Next Generation (IllustrisTNG) project (\citealt{nelson2018first, pillepich2018first, naiman2018first, marinacci2018first, springel2018first}) is an upgraded version of the original 'Illustris' project (\citealt{genel2014introducing, vogelsberger2014introducing, nelson2015illustris}) and it is composed by 18 simulations, 9 of them are magneto-hydrodynamical running within 3 different volumes of 50, 100 and 300 Mpc per side (named TNG50, TNG100 and TNG300, respectively), each of them at three different resolution levels evolving from z=$127$ to z=$0$. The simulations are run with the moving-mesh AREPO code (\citealt{springel2010}) under a quasi-Lagrangian, second-order treatment in space and time. The simulation uses parameters from Planck Cosmology (\citealt{Plack2016planck}) where the average densities of baryonic matter, dark energy, and matter are $\Omega_b = 0.0486$, $\Omega_\Lambda = 0.0.6911$ and $\Omega_m =  0.3089$ and a Hubble-Lemaitre parameter value of $h= 0.6774$. IllustrisTNG uses a Friends-of-Friends (FoF) algorithm with linking length $l=0.2$ (\citealt{davis1985evolution}) (in units of the mean particle separation), within which gravitationally bound substructures (galaxies) are located and hierarchically characterised; particles below this threshold are not considered associated (\citealt{pillepich2018simulating}). The \texttt{SUBFIND} algorithm (\citealt{springel2001populating}) identifies gravitationally bound substructures (galaxies), which are interpreted as substructures inside FoF halos. 

In the simulation, the central galaxy is usually identified as the most massive galaxy at the center of the FoF structure, while satellite galaxies are the remainder of galaxies in the FoF haloes. Inside a FoF halo exists one central galaxy and multiple satellites. 

In this paper, we employ the TNG100 simulation because it offers a good balance between its cosmological volume and spatial resolution (with a softening length below one kpc). This simulation contains $100$ snapshots in a range from $z=20$ to $z=0$. We use snapshot $99$ which corresponds to a redshift $z=0$. This simulation follows the evolution of $2 \times 1830^3$ particles that include dark matter (DM)  and gas cells. The initial mass  is $7.49 \times 10^6$ M$_{\odot}$  and $1.39 \times 10^6$ M$_{\odot}$ for DM particles and gas cells, respectively. The spatial resolution is $0.74$ kpc, determined by the gravitational softening length of DM and stellar particles. 
 
\subsection{Sample construction}

In our analysis, we focus on galaxies with a stellar mass above M$_{\star} \geq 10^{10}$ M${\odot}$, ensuring that we have well-resolved discs consistent with the simulation resolution, as described in \cite{zhao2020barred}. This previous work serves as the basis for segregating galaxies with and without stellar bars which will be explained later in this section. Furthermore, we specifically focus on disc-dominated galaxies for which we adopt the same classification criteria presented by \cite{perez2022formation}, which employs the $\kappa_{\mathrm{rot}}$ parameter defined by  (\citealt{sales2010feedback}) as

\begin{equation}
\centering
    \kappa_{\mathrm{rot}}= \frac{K_{\mathrm{rot}}}{K}= \frac{1}{K} \sum \frac{1}{2} m_i \left( \frac{j_{z,i}}{R_i} \right)^2 
\end{equation}

where $K_{\mathrm{rot}}$ corresponds to the kinetic energy of the stellar component along the azimuthal component, and $K$ is the total kinetic energy. In the fullest expression, the sum is overall stellar particles in the galaxy, $m_i$ is the mass of a particle, $j_{z, i}$ is the specific angular momentum of the z-component, and $R_{i}$ is the projected radius. For the purpose of classification, if $\kappa_{\mathrm{rot}}$ value is $ \geq 0.5$ (\citealt{perez2022formation}), the galaxy is considered a late-type, otherwise we consider it to be an early-type galaxy. Using this stellar mass and morphology selections, our parent sample comprises $4, 224$ disc-dominated galaxies from TNG100. 

Once the disc-dominated galaxies have been identified, we proceed to segregate the sample between LSBs and HSBs. As mentioned in the introduction, we will use the value of their central surface brightness in the $r-$band ($\mu_r$) calculated by \cite{perez2022formation}, where they compute $\mu_r$ following \cite{zhong2008large} and \cite{bakos2012deep} for all galaxies with stellar mass M$_{\star} > 10^9$ M$_{\odot}$ in TNG100 at z=0. The central surface brightness parameter is defined as:

\begin{equation}
\centering
    \mu_r= m_r + 2.5 \log (\pi r^{2}_{50,r})
\end{equation}

where $m_r$ is the apparent magnitude inside $r_{50,r}$ which is the projected half-light radius. Their whole sample of galaxies was analyzed in a face-on view such that the angular momentum vector for the stellar component is orientated along the line of sight. We do not consider dust attenuation effects in  $\mu_r$ values since it has no significant impact in face-on view galaxies as reported by \cite{kulier2020massive}. Using this identification in our total sample of $4, 224$ galaxies we find two subsamples:  $3,572$ HSBs (HSBs), and $652$ LSBs. From \cite{perez2022formation} we can also identify how many central galaxies and satellites there are for each subsample. We find that $1, 772$ and $489$ HSBs and LSBs are central galaxies, respectively, and $1,800$ and $163$ satellite galaxies for HSBs and LSBs.  

Finally, in order to classify barred and unbarred galaxies, we adopt the \cite{zhao2020barred} catalogue, which uses the ellipse fitting method that measures ellipticity ($\epsilon$) and position angle (PA) by analyzing the elliptical isodensity contours of face-on mass surface density maps. For a bar, they use the identification criteria from \cite{marinova2007characterizing}, where within the bar, the maximum value of the ellipticity must be $\epsilon > 0.25$, with a decrease of $>0.1$ outwards, and a variation of the PA  less than $10^{\circ}$.  In addition to this identification, \cite{zhao2020barred} performed a visual inspection of the images to verify the morphologies of barred galaxies. Of the $1,172$ identified bars, $853$ ($\sim 73\%$) correspond to strong bars with a value of $\epsilon \geq 0.4$, while the remaining $319$ ($\sim 27\%$) are considered weak bars with $0.25< \epsilon < 0.4$. Even though the results presented in the following sections correspond to the total bar fraction (including both strong and weak bars), the quantitative and qualitative behaviors are primarily dominated by the strong bars, which constitute the majority and where the identification is more reliable. The barred galaxies in our sample exhibit a wide range of bar strengths, with $\epsilon$ varying between 0.25 and 0.8, independent of stellar mass, consistent with observations from NIR surveys (\citealt{diaz2016characterization}). In contrast, the distribution of bar sizes does show a dependence on stellar mass, with more massive galaxies hosting larger bars, a trend also reproduced in our sample and consistent with \cite{diaz2016characterization} as discussed in \cite{zhao2020barred} (see their Figure 4). Due to resolution limitations, galaxies with bar sizes smaller than 1.4 kpc are underrepresented in the catalogue, therefore, in this work, we only consider bars with sizes greater than 1.4 kpc. 

In our parent sample, galaxies are categorized as barred only if listed in the \cite{zhao2020barred} catalogue; otherwise, they are considered unbarred, they fall within the stellar mass range of M$_{\star} \geq 10^{10}$ M${\odot}$ and are predominantly rotation-supported ($\kappa_{\mathrm{rot}} \geq 0.5$). In our total sample of 4,224 galaxies, there are 1,172 classified as barred and 3,052 as unbarred based on these criteria.  Additionally, we ran tests using the catalogue of \cite{rosas2020buildup}, which includes over $300$ massive galaxies ($10^{10.4-11} M_{\odot}$) with well-defined cold discs and stable stellar bars, ensuring their persistence in two consecutive snapshots. These galaxies were identified by Fourier decomposing the face-on stellar mass density. Both catalogues demonstrate comparable bar fractions in relation to the stellar mass. Despite this, we chose to use the \cite{zhao2020barred} catalogue due to its enlarged statistical sample and wider stellar mass range. 

\begin{figure*}
\captionsetup[subfigure]{labelformat=empty}
\begin{subfigure}{0.4\linewidth}
    \includegraphics[width=\linewidth, height=0.9\linewidth]{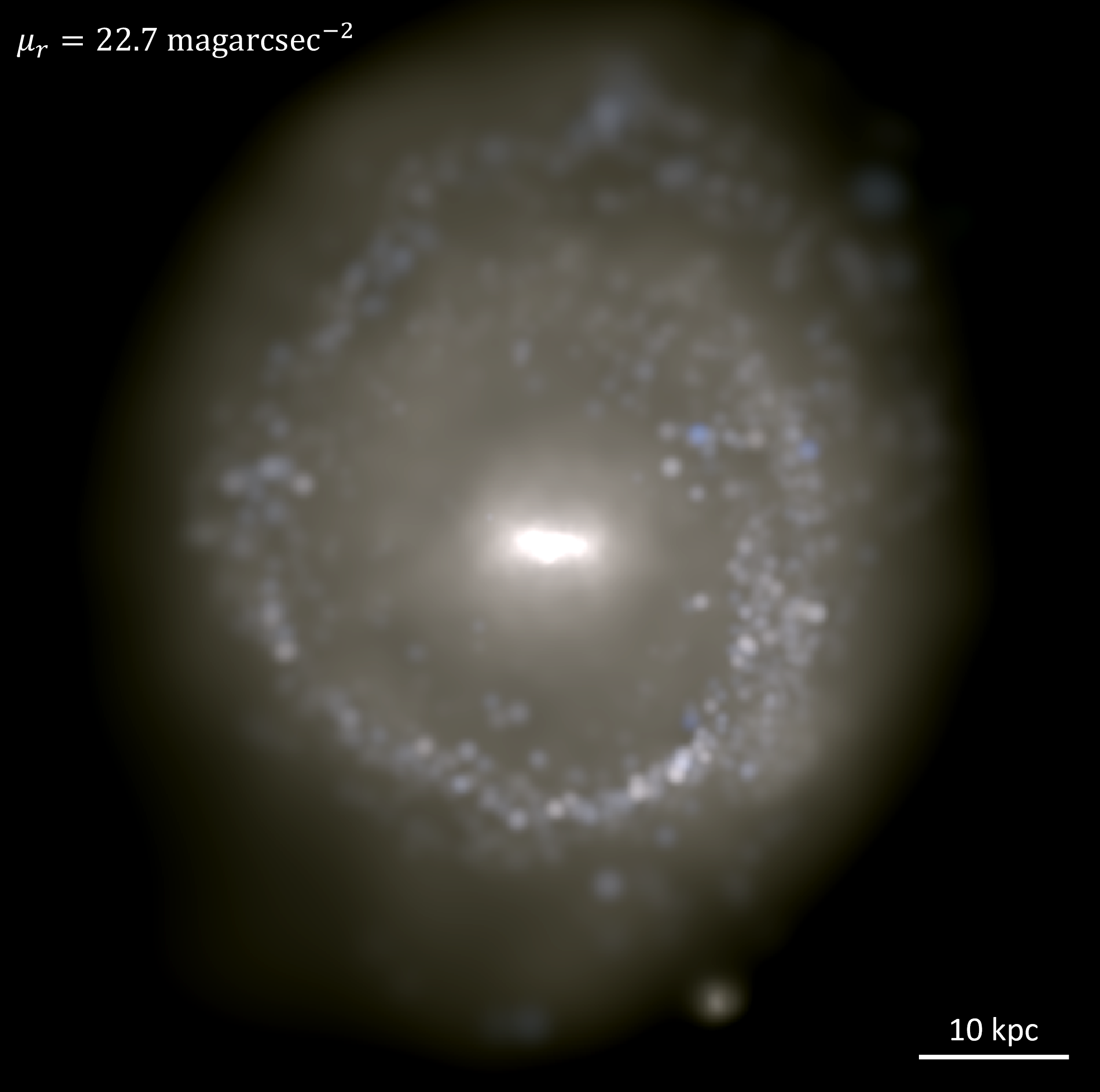}
    \subcaption{  }
\end{subfigure}
\begin{subfigure}{0.4\linewidth}
    \includegraphics[width=\linewidth, height=0.9\linewidth]{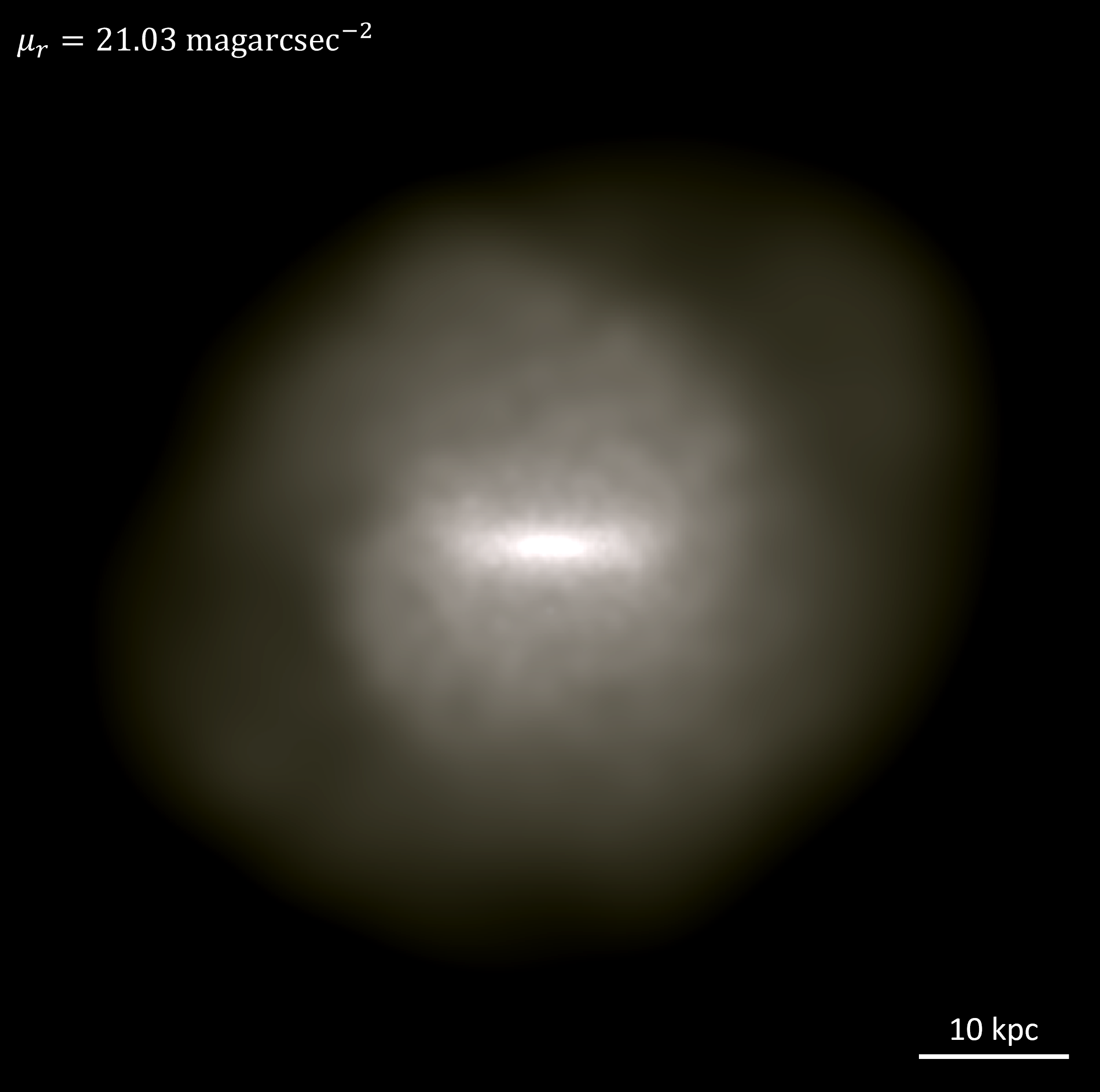}
  \caption{  }
\end{subfigure}

\begin{subfigure}{0.4\linewidth}
    \includegraphics[width=\linewidth, height=0.9\linewidth]{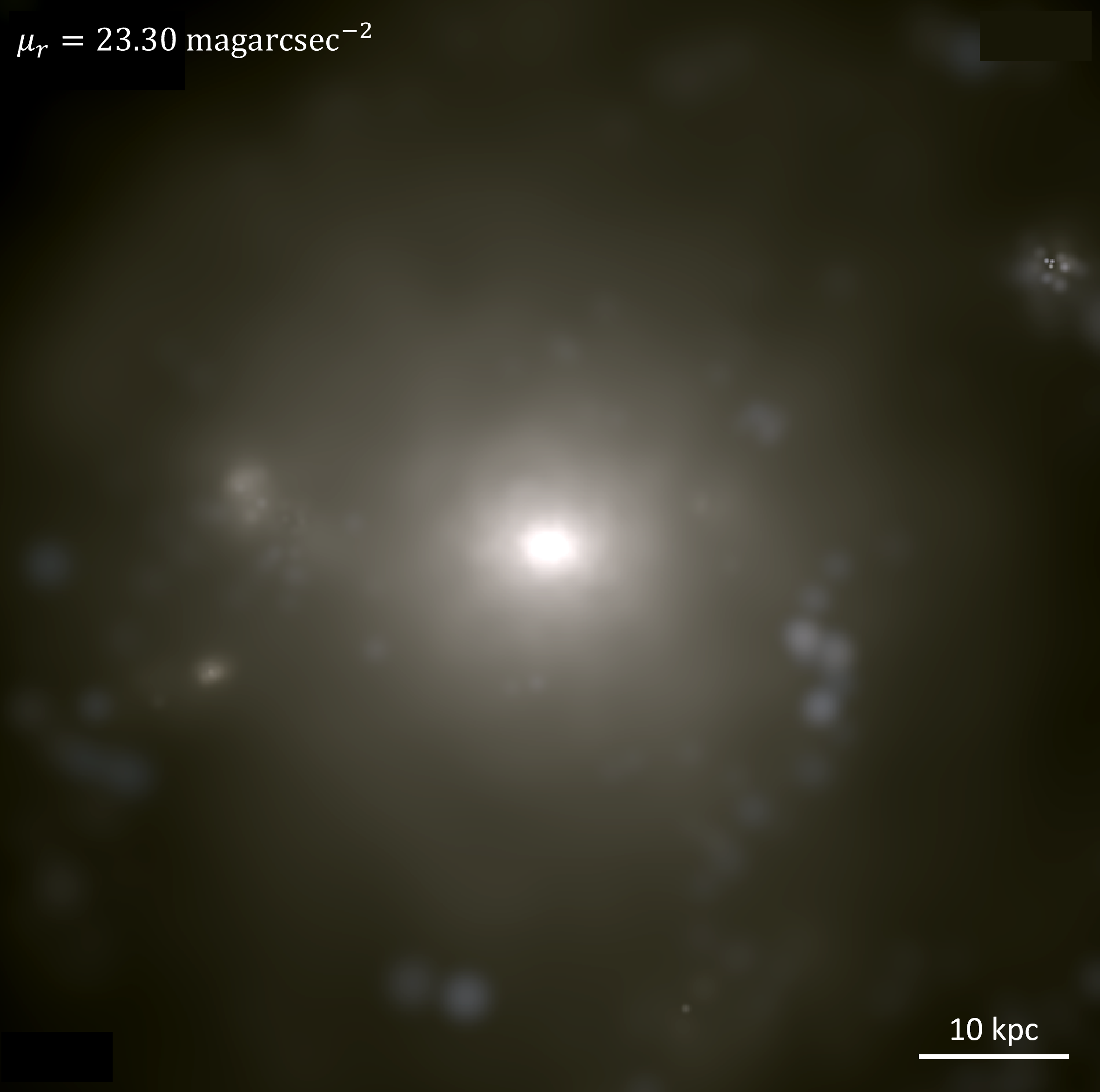}
  \caption{  }
\end{subfigure}
\begin{subfigure}{0.4\linewidth}
    \includegraphics[width=\linewidth, height=0.9\linewidth]{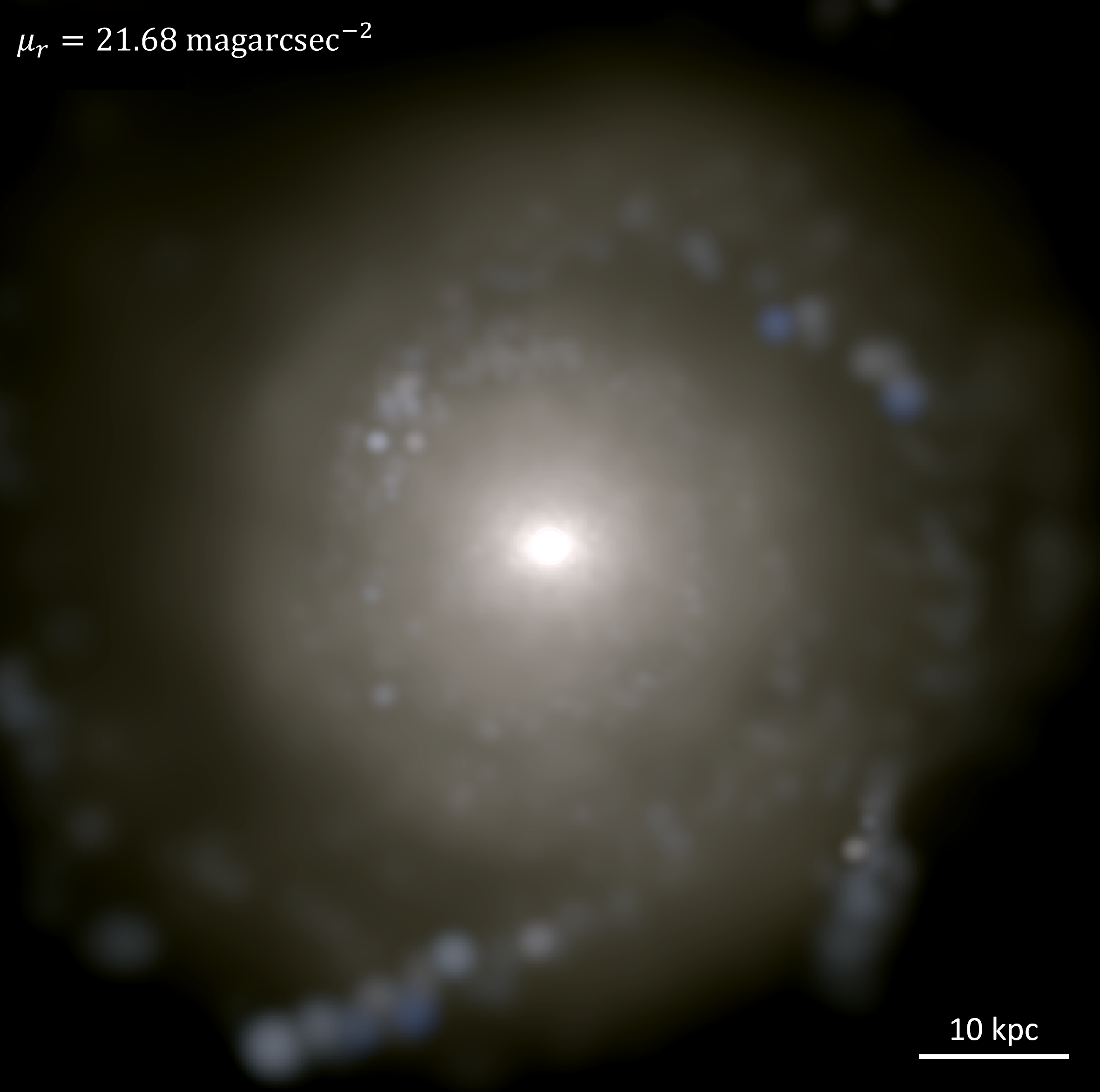}
  \caption{  }
\end{subfigure}
\caption{Comparison between randomly selected barred (top panels) and unbarred (bottom panels), LSB (left panels), and HSB (right panels) galaxies. We include on the upper left corner of each panel the values of the effective surface brightness of each system, all of them with stellar mass $ \sim 10^{10.9}$ M$_{\odot}$.}
\label{galcomp}
\end{figure*}

In Fig. \ref{galcomp} we present examples of both LSB (top left) and HSB (top right) barred galaxies, along with unbarred LSB (bottom left) and HSB (bottom right) counterparts. These galaxies have similar stellar masses, approximately $10^{10.9}$ M$_{\odot}$. As expected, the LSBs appear more spatially extended than the HSBs in both barred and unbarred examples. In each of the top panels, a noticeable bar structure is present in the center of the galaxy, while in the lower panels, the elongated structure of the bar is absent. The whole sample composition is present in Table \ref{comp-sample}. 

\begin{table}
\caption{Composition of the constructed galaxy sample, segregated according to their surface brightness ($\mu_r$), the presence of a stellar bar, and whether they are considered central or satellite galaxies.} 
\centering
\begin{tabular}{c c c}
\hline\hline
& Total Sample: \textbf{4,224} galaxies & \\
\hline\hline
& HSBs: \textbf{3,572} & LSBs: \textbf{652} \\
\hline
Barred & $1,011$ ($28.11 \%$) & $161$ ($24.73 \%$) \\
Non-barred & $2,561$ ($71.89 \%$) & $491$ ($75.27 \%$) \\
\hline
Central & $1,772$ & $489$ \\
Satellite & $1,800$ & $163$ \\[1ex]
\hline
\end{tabular}
\label{comp-sample}
\end{table}

\subsection{\textbf{Galactic Parameters of the sample.}}
\label{physicalpar}

The physical parameters on which we focus our analysis include the gas mass fraction, $f_{\mathrm{gas}}$, the bulge-to-total mass fraction, B/T, and the spin parameter of the host halo, $\lambda$. In addition to these parameters, we consider the tidal parameter $Q$ to characterise the local environment. In the following subsections, we describe the methods employed to calculate each. 

\subsubsection{Gas mass fraction.}
The gas mass fraction $f_{\mathrm{gas}}$ is calculated within twice the effective radius, and defined as the ratio between gas mass content and baryonic mass considering stellar and gas mass, i.e:

\begin{equation}
\label{eqgas}
    f_{\mathrm{gas}}= \dfrac{\mathrm{M}_{\mathrm{gas}}}{\mathrm{M}_{\mathrm{gas}}+ \mathrm{M}_{\star}}
\end{equation}
\\

If we instead take into account the atomic and molecular hydrogen masses estimated by \cite{Diemer18}, following \cite{springel2003cosmological}, our findings remain consistent. This is because the total atomic and molecular hydrogen gas is primarily contained within the two effective radii.
The $f_{\mathrm{gas}}$ distribution for samples HSBs and LSBs is presented in Fig.~\ref{fig:sfig1}, where median values are indicated by down pointing arrows, while the 25th and 75th percentiles are represented by dotted lines. All parameter distributions in Fig. \ref{pardist} follow the same style. For the gas fraction distribution (Fig.~\ref{fig:sfig1}) we can see that both subsamples show the same range of $f_{\rm \mathrm{gas}}$, but HSBs are concentrated at low f$_{\mathrm{gas}}$ values. The majority of HSBs have gas fractions $<0.2$. On the other hand, LSBs are more uniformly distributed over the entire range of values. The median gas mass fractions for HSBs and LSBs are $0.122$ and $0.207$, respectively, showing that, on average,  LSBs are more gas-rich than their HSBs counterparts. This result is in good agreement with observational studies (\citealt{o2004new, huang2014highmass}), as well as with results from cosmological simulations (\citealt{di2019nihao}, \citealt{perez2022formation}). 

\subsubsection{Bulge-to-total mass fraction}

Although our sample construction is based on the $\kappa_{\mathrm{rot}}$ parameter to select systems where the stellar kinetic energy invested in the rotation is dominant (as required by the \citealt{zhao2020barred} and \citealt{perez2022formation} catalogues), this parameter does not directly translate to any particular morphological criterion employed in observational studies. 
To account for the morphology of the simulated galaxies, we turn to the bulge-to-total mass fraction, as computed in the \cite{genel2015galactic} catalogue. To distinguish between the morphological components of the galaxies in TNG100, the authors utilized the circularity parameter $\eta = J_z/J(E)$, where $J_z$ is the specific angular momentum of each stellar particle and $J(E)$ is the maximum angular momentum found among the stellar particles, determined from a list sorted by binding energy. To mitigate numerical fluctuations, the maximum value of $J(E)$ is computed as the average over a range of 50 particles before and after the peak value in the sorted list. As described in \cite{genel2015galactic}, twice the fractional mass of stars with $\eta < 0$ represents the bulge component. This approach assumes that the bulge is primarily non-rotating, although some galaxies may contain rotating bulges. 

The $B/T$ distribution for our LSB and HSB subsamples are presented in Fig.~\ref{fig:sfig2}, showing that both follow similar distributions with low $B/T$ values, mostly restricted to values below $B/T <0.6$. Thus, the bulge component in both subsamples is not massive in comparison to the total mass in the galaxy. There is no significant difference in the median $B/T$ value. However, LSBs have a slightly smaller median $B/T$ ($0.238$) than $B/T$ in HSBs ($0.257$). 

\subsubsection{Spin parameter}
To calculate the spin parameter denoted by $\lambda$, we use the expression proposed by \cite{bullock2001profiles} that relates the total angular momentum of halo $J$, the virial radius $R$, virial mass $M$ and virial circular velocity $V$:

\begin{equation}
\label{lambda}
    \lambda= \dfrac{J}{\sqrt{2} MRV}
\end{equation}

It is important to point out that the spin of the dark matter halo is well-defined exclusively on central galaxies, which limits our study to only these kinds of systems. In our sample, there are only $2,261$ central galaxies ($53$\% of the total sample), of which $489$ are LSBs and $1,772$ are HSBs. The distribution for the adimensional spin parameter is shown in Fig.~\ref{fig:sfig3}. HSBs present smaller values of the spin parameter than their LSB counterparts. Most HSBs have values $\lambda <0.08$ while the distribution for LSBs is shifted to larger values. Median values are $0.043$ and $0.070$ for HSBs and LSBs, respectively. 

\subsubsection{Environment and the tidal parameter.}
\begin{figure*}
\captionsetup[subfigure]{labelformat=empty}
\begin{subfigure}{0.48\linewidth}
    \includegraphics[width=\linewidth, height=0.7\linewidth]{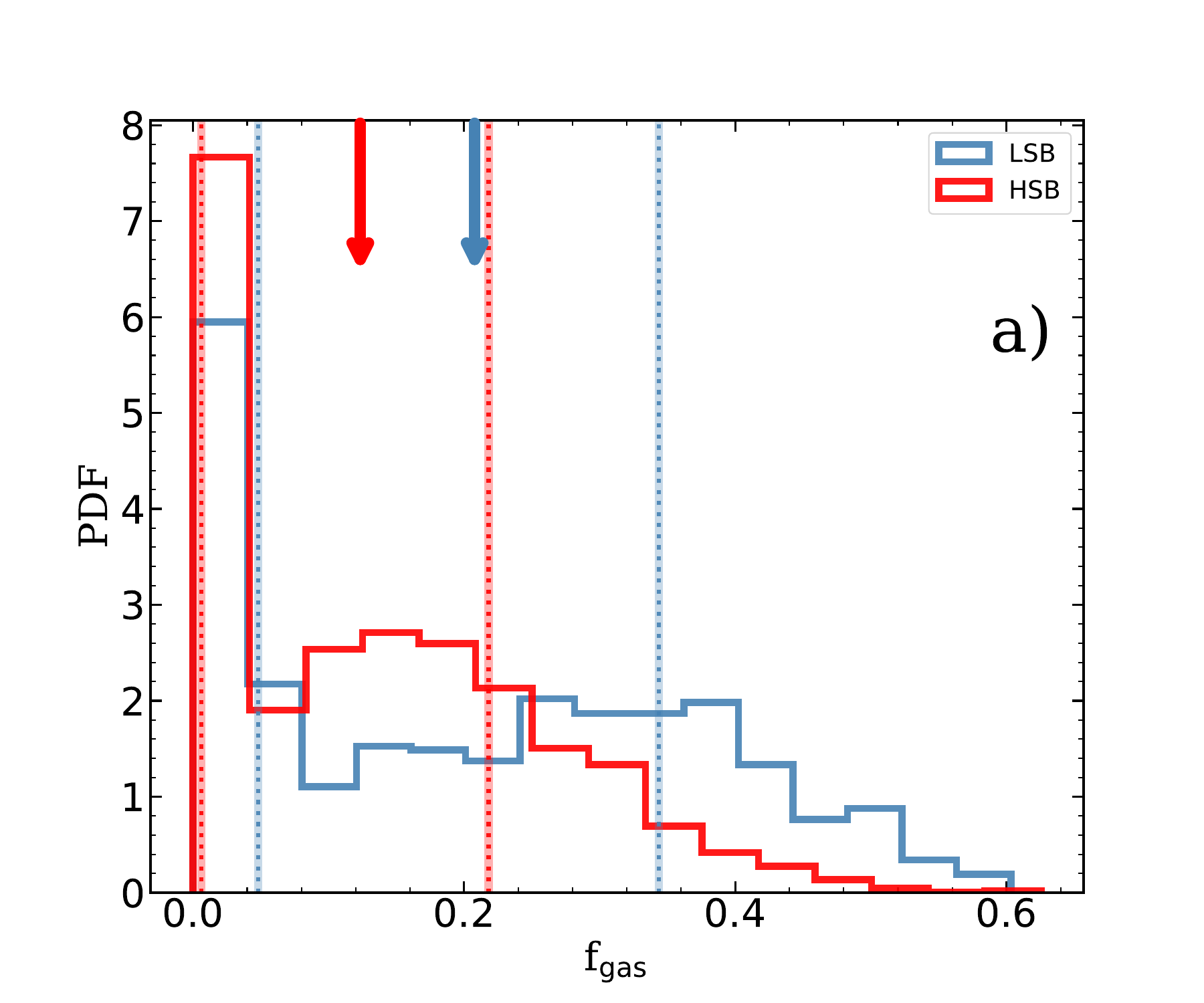}
    \subcaption{  }
    \label{fig:sfig1}
\end{subfigure}
    \hfill
\begin{subfigure}{0.48\linewidth}
    \includegraphics[width=\linewidth, height=0.7\linewidth]{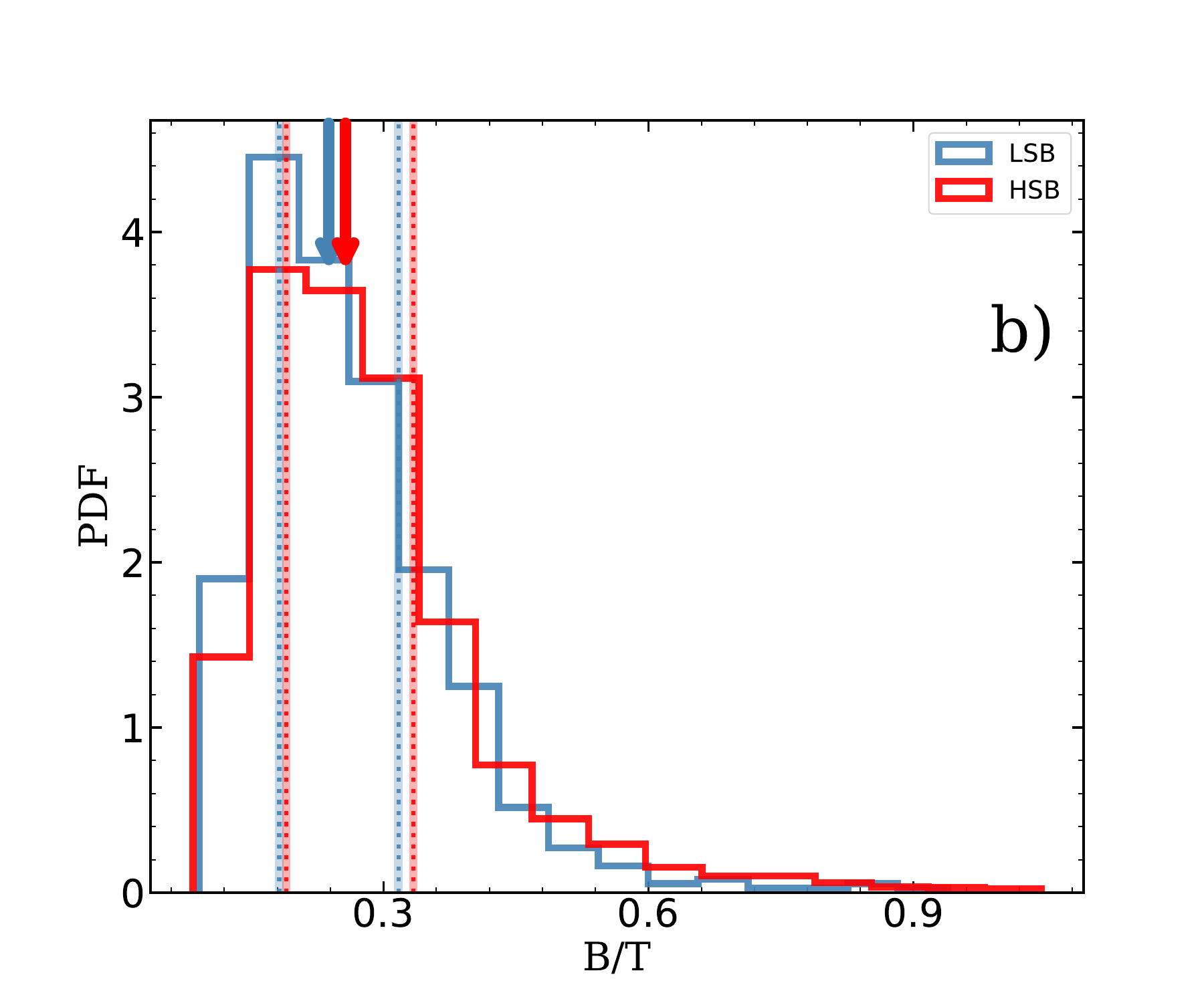}
  \caption{  }
  \label{fig:sfig2}
\end{subfigure}

\begin{subfigure}{0.48\linewidth}
    \includegraphics[width=\linewidth, height=0.7\linewidth]{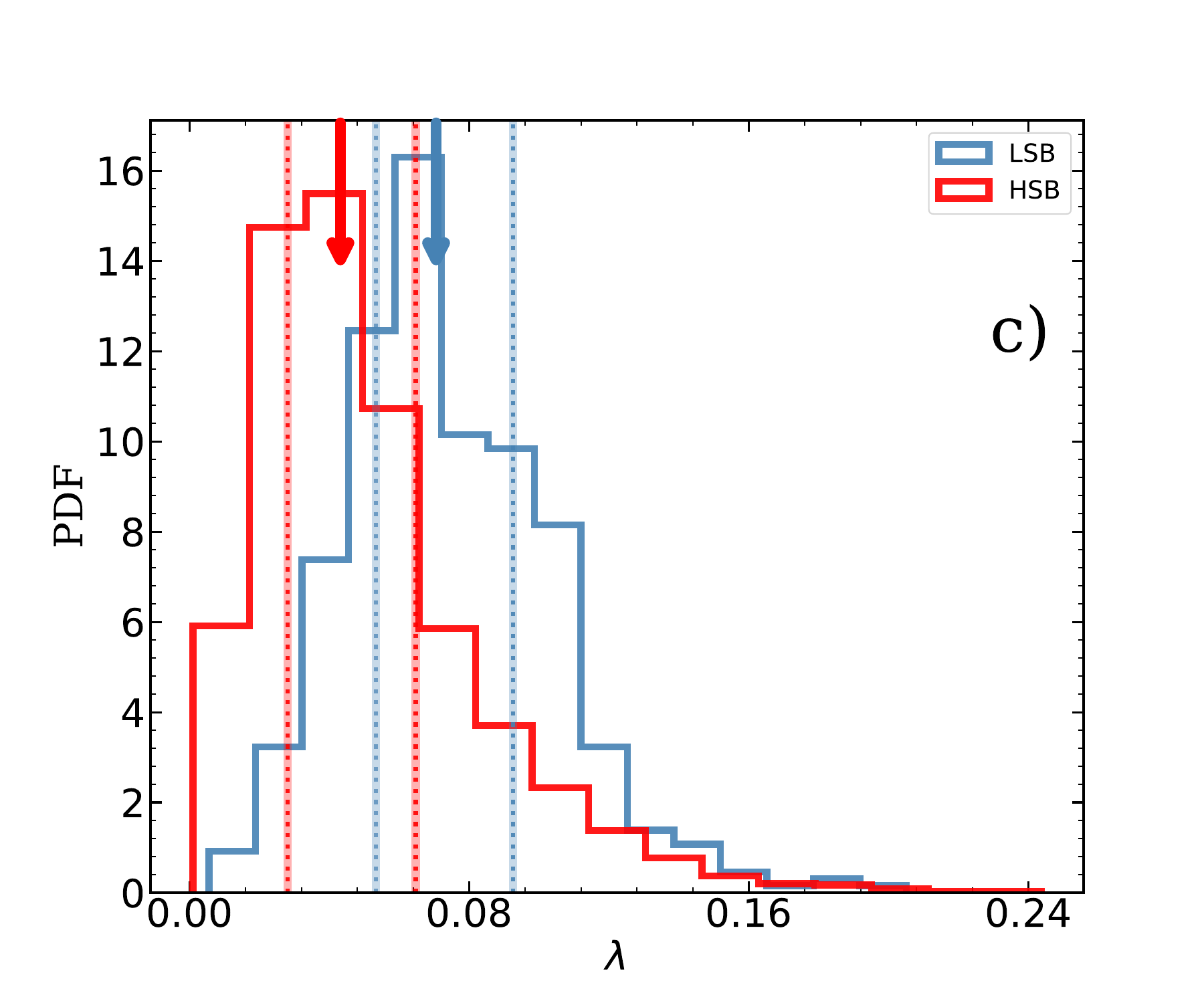}
  \caption{  }
  \label{fig:sfig3}
\end{subfigure}
    \hfill
\begin{subfigure}{0.48\linewidth}
    \includegraphics[width=\linewidth, height=0.7\linewidth]{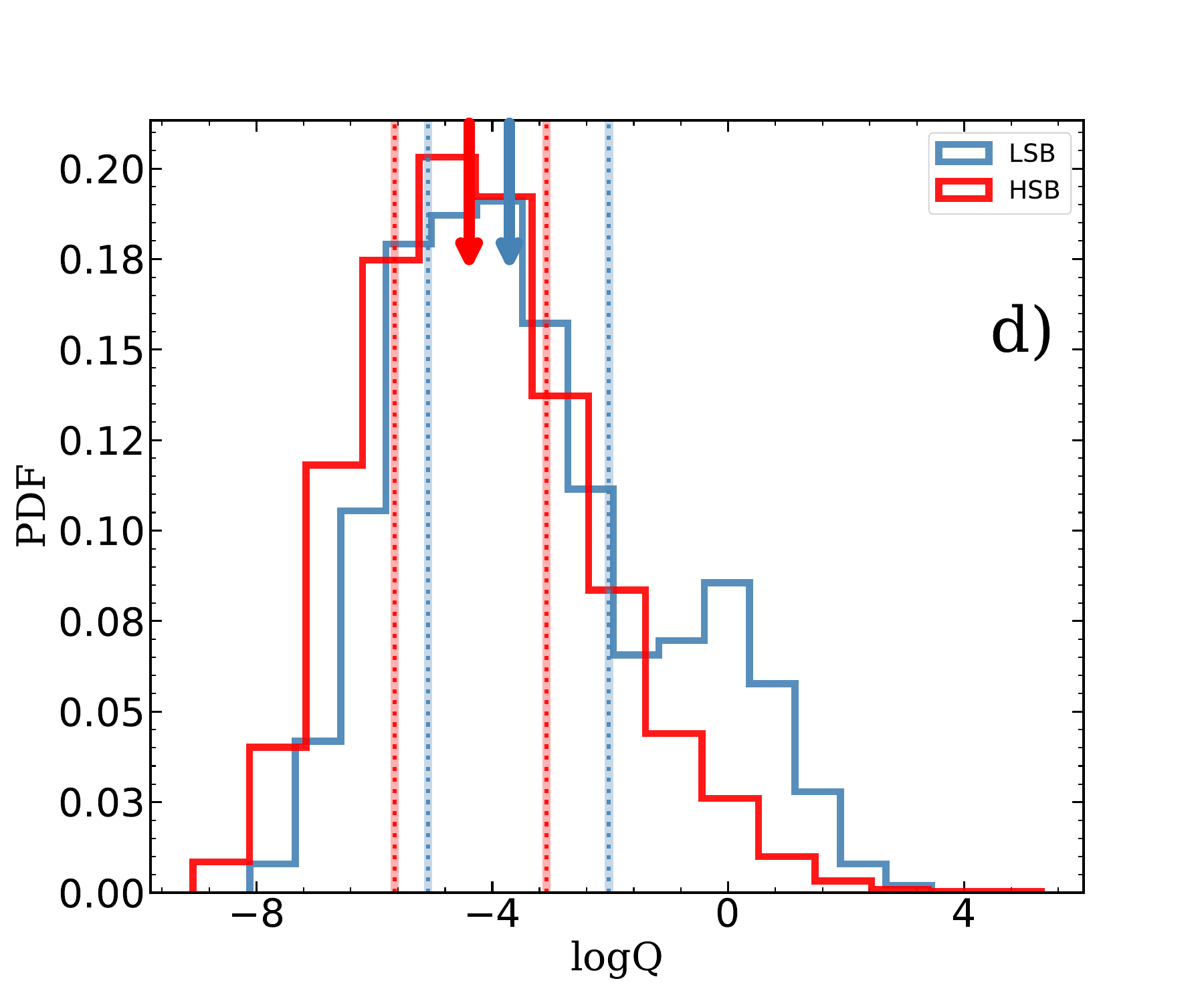}
  \caption{  }
  \label{fig:sfig4}
\end{subfigure}
\caption{(a) Gas mass fraction, (b) bulge-to-total mass fraction, (c) spin parameter (only for centrals), and (d) tidal parameter distributions for LSBs (blue lines) and HSBs (red lines) samples. The down pointing arrows represent the median values of each sample. The percentiles $25^{\rm th}$ and $75^{\rm th}$ of the distributions are represented by the dotted lines. Notably, LSBs exhibit higher values for gas mass fraction, spin, and tidal parameter compared to their HSB counterparts. The distributions for the bulge-to-total mass fraction are similar for both galaxy types.} 
\label{pardist}
\end{figure*}

Finally, to analyze how the environment affects the bar fraction in our sample, we characterised the environment by employing the tidal parameter $Q$. This dimensionless quantity measures the strength of interaction between a galaxy and its companion and it is defined by \cite{dahari1984companions} as:

\begin{equation}
    Q= \dfrac{M_i}{M_\mathrm{p}} \cdot \left(\dfrac{D_\mathrm{p}}{S_{\mathrm{ip}}}   \right)^3 
\label{eqQ}
\end{equation}

where $M_\mathrm{p}$ is the stellar mass of a given galaxy with diameter $D_\mathrm{p}$ defined as 10 times the stellar effective radius, $M_i$ is the stellar mass of its closest neighbor located at a distance $S_{ip}$. As we can see from Eq. \ref{eqQ}, $Q$ has a positive relation with galaxy diameter but a negative one with the distance. As contrast to HSBs, LSBs are more extended and have less dense dark matter halos (\citealt{mcgaugh2001high}). In addition, LSBs have a lower surface stellar density. These characteristics make them more susceptible to tidal forces from their neighbors, then it is expected that they have higher $Q$ values than HSBs, which are more compact and may not be as affected by their companions. Tidal parameter distribution for subsamples is present in Fig.~\ref{fig:sfig4} where we can see that the median value for $Q$ is slightly higher for LSBs than HSBs with $\log Q= -3.76$ and $-4.43$, respectively. 

It is worth mentioning that the distributions of gas mass fraction, bulge-to-total mass fraction, and tidal parameter shown in Fig. \ref{pardist} remain essentially unchanged when considering only central galaxies. This indicates that including or excluding satellites does not introduce significant bias in these distributions. In particular, the median values of gas fraction and bulge-to-total mass fraction show no substantial differences. For the tidal parameter, the difference between LSBs and HSBs becomes more pronounced when only central galaxies are considered, with LSBs maintaining a higher median value than HSBs. These findings support the robustness of the trends shown in Fig. \ref{pardist}, allowing for a consistent comparison between panels even when using only centrals in the $\lambda$ distribution.

\subsection{Control Samples}
\label{CS-desc} 
To control the effects of each selected parameter on the bar fraction, we constructed several control samples (CS). The control samples were based on the sample of LSBs. Using the \texttt{R} package \texttt{MatchIT}, we select for each LSB galaxy a set of HSBs to which different statistical weights are assigned such that the parameter distributions between LSBs and HSBs are statistically similar. \texttt{MatchIT} offers an easy-to-use interface for a range of matching methods that enhance the balance of covariates in observational studies, in order to reduce confounding and reliance on specific models when estimating treatment effects. The similarity of the distributions of the selected parameters in the control samples was corroborated by a Kolmogorov-Smirnov (KS) test. \\ 

In our control samples, we consider the physical parameters mentioned previously (stellar mass, gas mass fraction, bulge-to-total fraction, and spin),  and the tidal force parameter, considered as external. Seven control samples were constructed, with the controlled parameters for each of them listed in Table \ref{tab1}. The first three control samples consider individual parameters in isolation. The following three control samples use pairs of parameters to determine if the combination of these parameters can explain the differences in bar fraction. Finally, CS7 takes into account the tidal parameter to analyze the environmental effects. The probability distribution functions (PDFs) for our control and uncontrolled samples are presented in the Appendix (section \ref{anexes}), where the p-values from the KS test are also shown.

\begin{table}
\caption{Control samples description.From left to right: Name and the parameter(s) that are controlled.} 
\centering      
\begin{tabular}{c c }  
\hline\hline                        
Control sample & Controlled parameters   \\ 
\hline  
CS1 & M$_{\star}$ \\   
CS2  & f$_{\mathrm{gas}}$ \\
CS3  &  $\lambda$ \\ 
CS4  &  f$_{\mathrm{gas}}$, $\lambda$ \\
CS5  &  $B/T$  \\
CS6  &  f$_{\mathrm{gas}}$, $B/T$ \\
CS7  &   $Q$ \\[1ex]
\hline     
\end{tabular} 
\label{tab1} 
\end{table}

\section{Results}
\label{results}
In this section, we present the bar fraction (ratio of the number of bars to the total number of galaxies) and its relation with the physical parameters of the galaxies associated with the presence of a stellar bar using the control samples described in section \ref{CS-desc}.

\subsection{Bar fraction}
\label{bf-sect}

We find that the bar fraction for our total sample ($4,224$ disc galaxies) is $27.78 \pm 0.67\%$. This fraction indicates that $1,172$ galaxies in our sample exhibit a bar structure. The bar fraction for both LSBs and HSBs samples is $24.73 \pm 1.73\%$ and $28.35 \pm 0.74\%$, respectively, with errors calculated using the bootstrapping resampling method and a $1\sigma$ confidence level. All error bars in the following figures were calculated using this method. In addition, in each bin, a minimum of $15$ galaxies was considered. When evaluating the statistical significance of the data, the global bar fractions in LSBs are slightly smaller than those in HSBs. While the disparity may be minimal, it becomes more pronounced when compared at fixed stellar mass, as it is the stellar mass one of the key physical properties identified as determining the formation and growth of bars.

\begin{figure}
\includegraphics[width=\columnwidth]{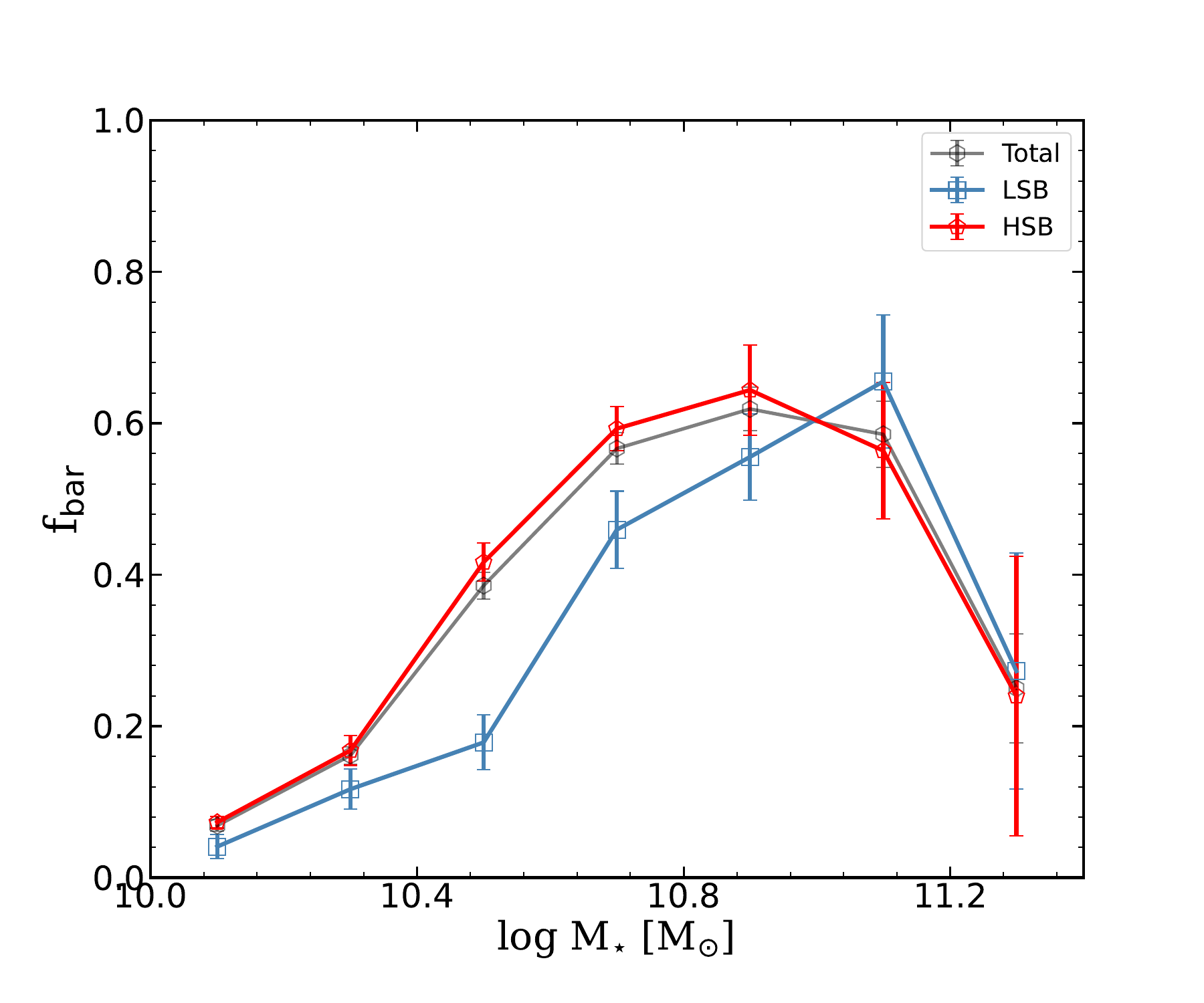} 
\caption{Bar fraction as a function of stellar mass for the parent galaxy sample (black line), replicating the findings presented by \protect\cite{zhao2020barred}. Blue and red lines correspond to the LSBs and HSBs samples, respectively. In all cases, the bar fraction increases towards the largest stellar mass values and then decreases. Across all cases, the bar fraction demonstrates an initial increase towards the highest stellar mass values, followed by a subsequent decrease. Error bars represent an interval of 1$\sigma$ calculated by the bootstrapping resampling method.}
\label{fbar-logM}
\end{figure}

In Fig. \ref{fbar-logM}, we present the bar fraction as a function of stellar mass. From the figure, we can see that the bar fraction increases with the stellar mass for the total sample (gray lines) and for both HSB and LSB subsamples represented by red and blue lines, respectively. For the stellar mass range M$_{\star}$= $10^{10}-10^{11}$ M$_{\odot}$ the total bar fraction rises from $\sim 0.1$ to $0.6$ before falling to $\sim 0.2$ for M$_{\star}>$ $10^{11}$ M$_{\odot}$. 

We observe that for M$_{\star} < 10^{11} M_{\odot}$, HSBs have a higher $f_{\rm bar}$ than LSBs, even when accounting for the error bars. This value increases from approximately $0.1$ to $0.6$ with increasing stellar mass. The bar fraction in LSBs also increases with M$_{\star}$, but it reaches a maximum bar fraction of approximately $0.5$ at M$_{\star}= 10^{11} M_{\odot}$. Notably, the difference in bar fraction for subsamples is most pronounced in the small and intermediate stellar mass range. In contrast, for galaxies in the higher mass range (M$_{\star} \geq 10^{11}$ M$_{\odot}$), there is a decrease in $f_{\rm bar}$, dropping from $0.5$ to approximately $0.2$. In this range, the disparity in  $f_{\rm bar}$ between LSBs and HSBs becomes less pronounced, collapsing at the highest mass bin.

\subsection{Bar fraction dependence with physical parameters}
\label{sect-fbar-par}

\begin{figure*}
\captionsetup[subfigure]{labelformat=empty}
\begin{subfigure}{0.45\linewidth}
    \includegraphics[width=\linewidth, height=0.85\linewidth]{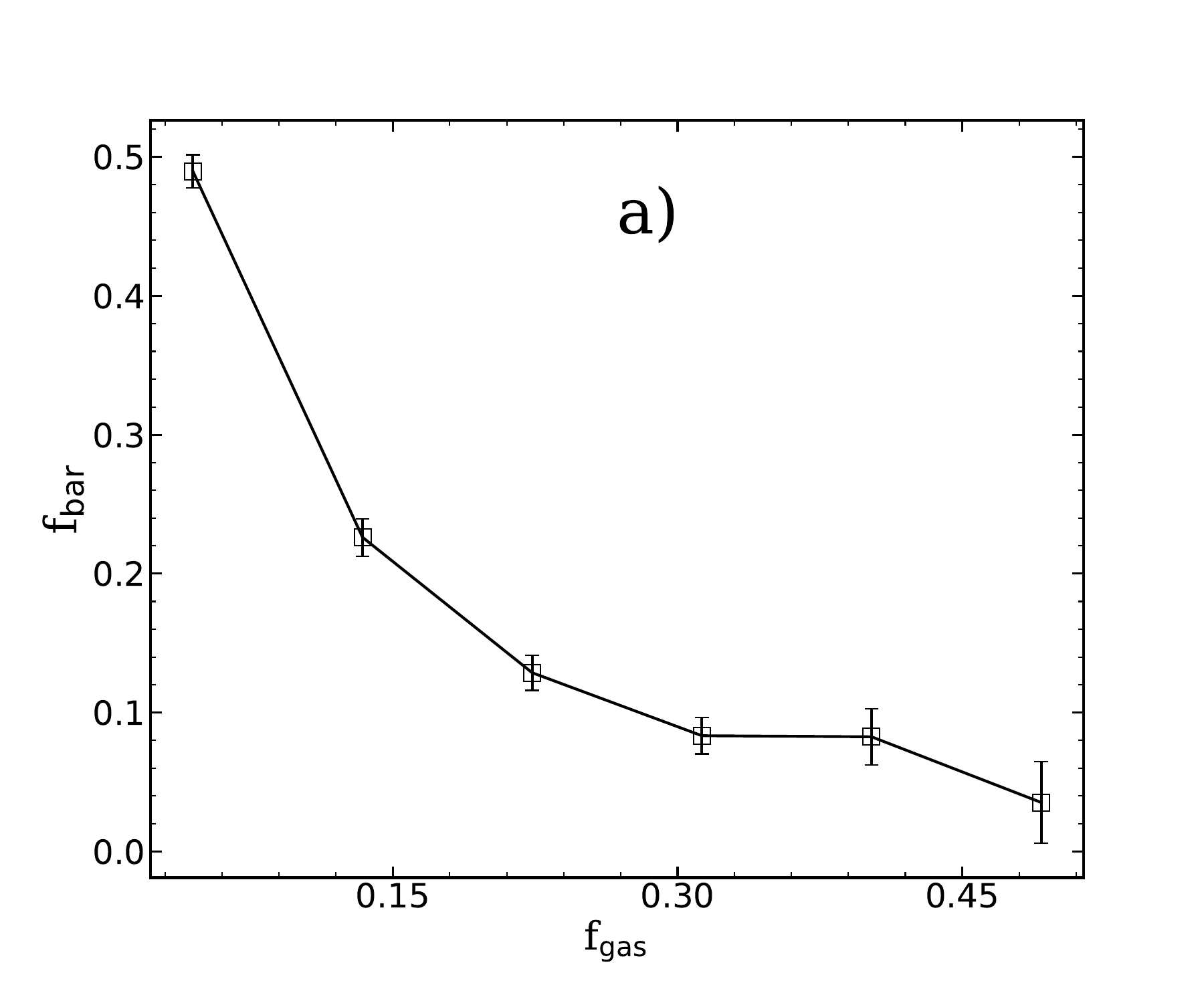}
\vspace{-0.6cm}
\caption{ }
    \label{totalfbarfgas} 
\end{subfigure}
    \hspace{.8cm}
\begin{subfigure}{0.45\linewidth}
    \includegraphics[width=\linewidth, height=0.85\linewidth]{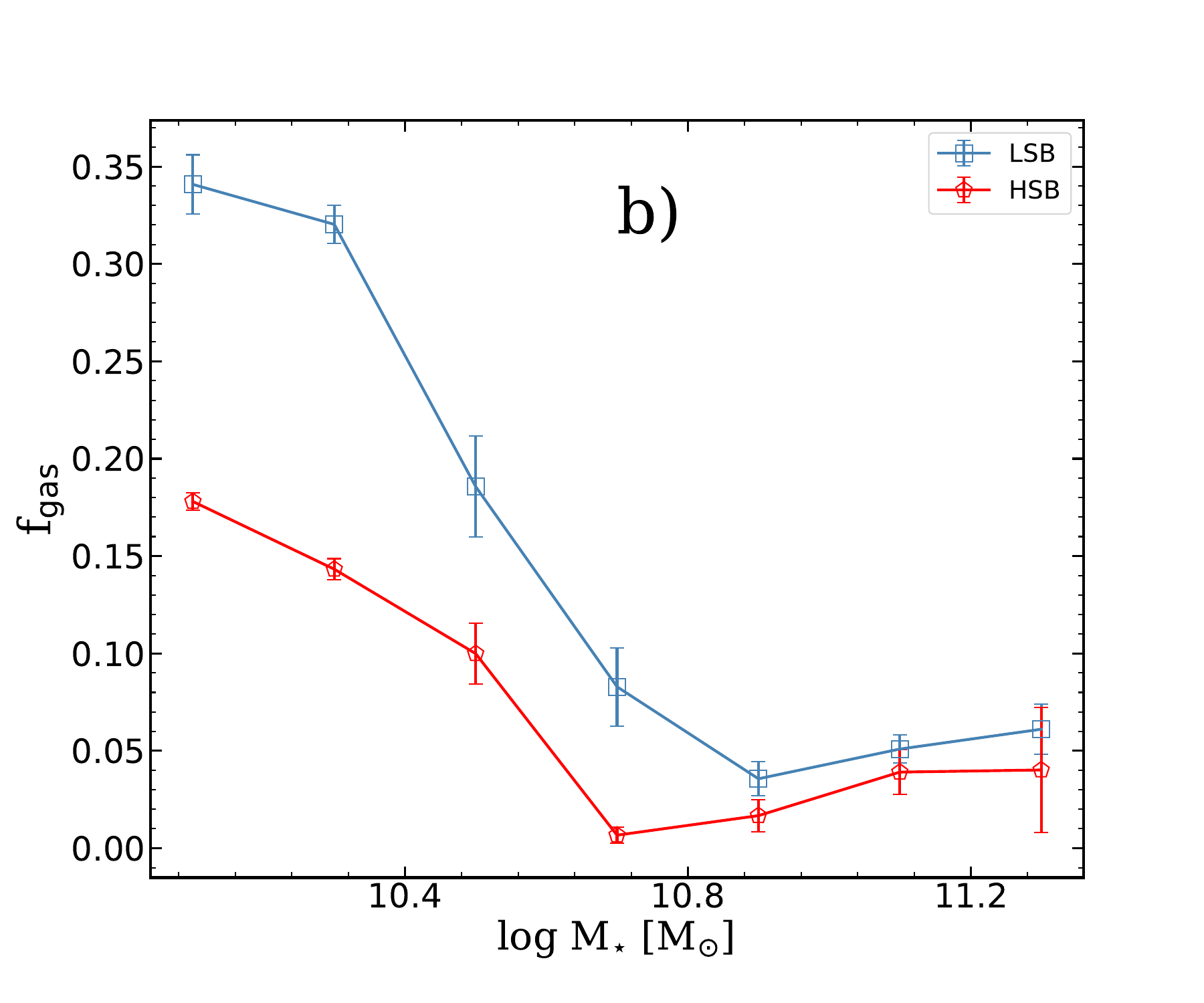}
\vspace{-0.6cm}
\caption{ }
    \label{subfgasMs}
\end{subfigure}

\vspace{-0.12cm}

\begin{subfigure}{0.45\linewidth}
    \includegraphics[width=\linewidth, height=0.85\linewidth]{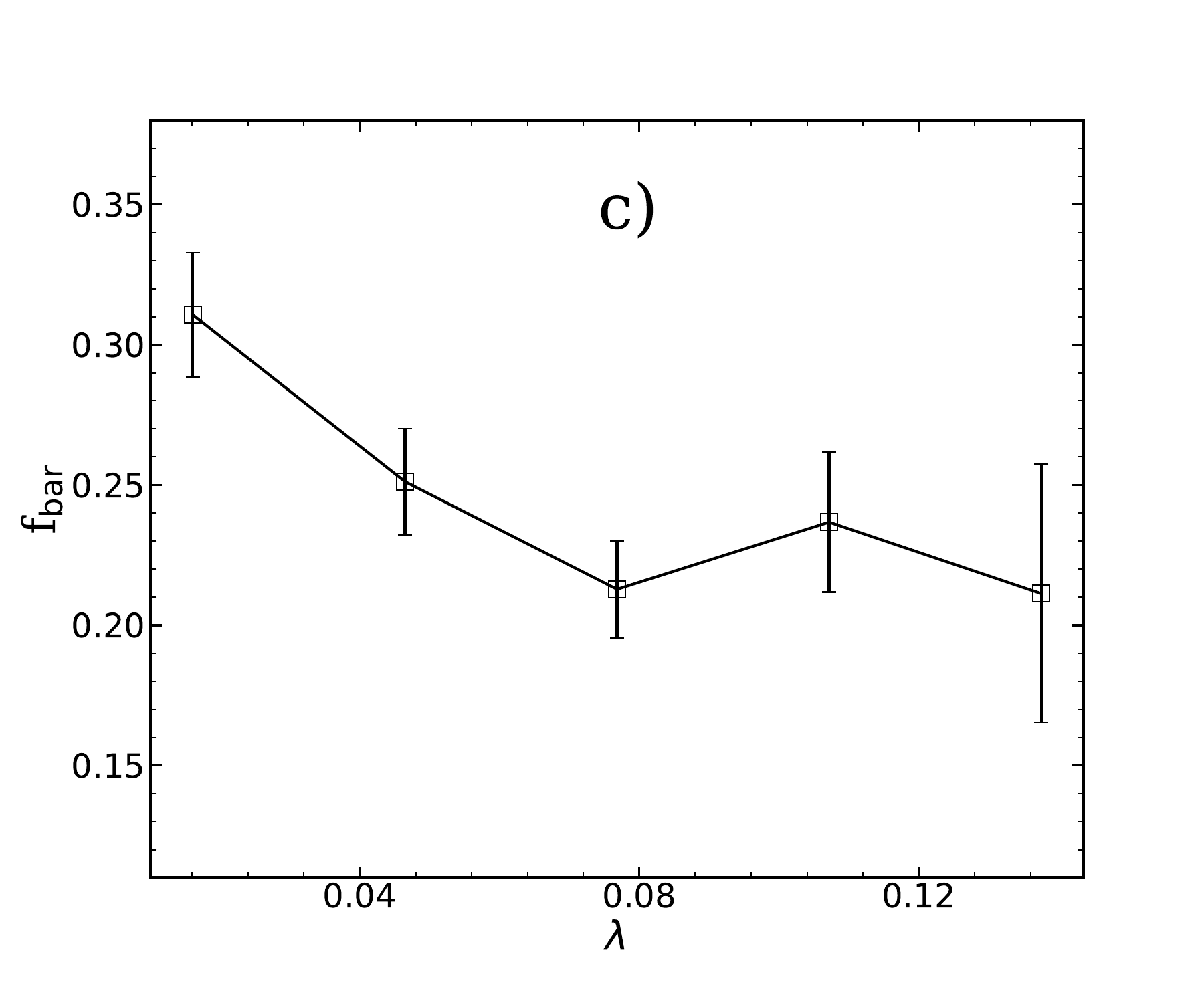} 
\vspace{-0.6cm}
\caption{ }
    \label{totalfbarlambda}
\end{subfigure}
    \hspace{.8cm}
\begin{subfigure}{0.45\linewidth}
    \includegraphics[width=\linewidth, height=0.85\linewidth]{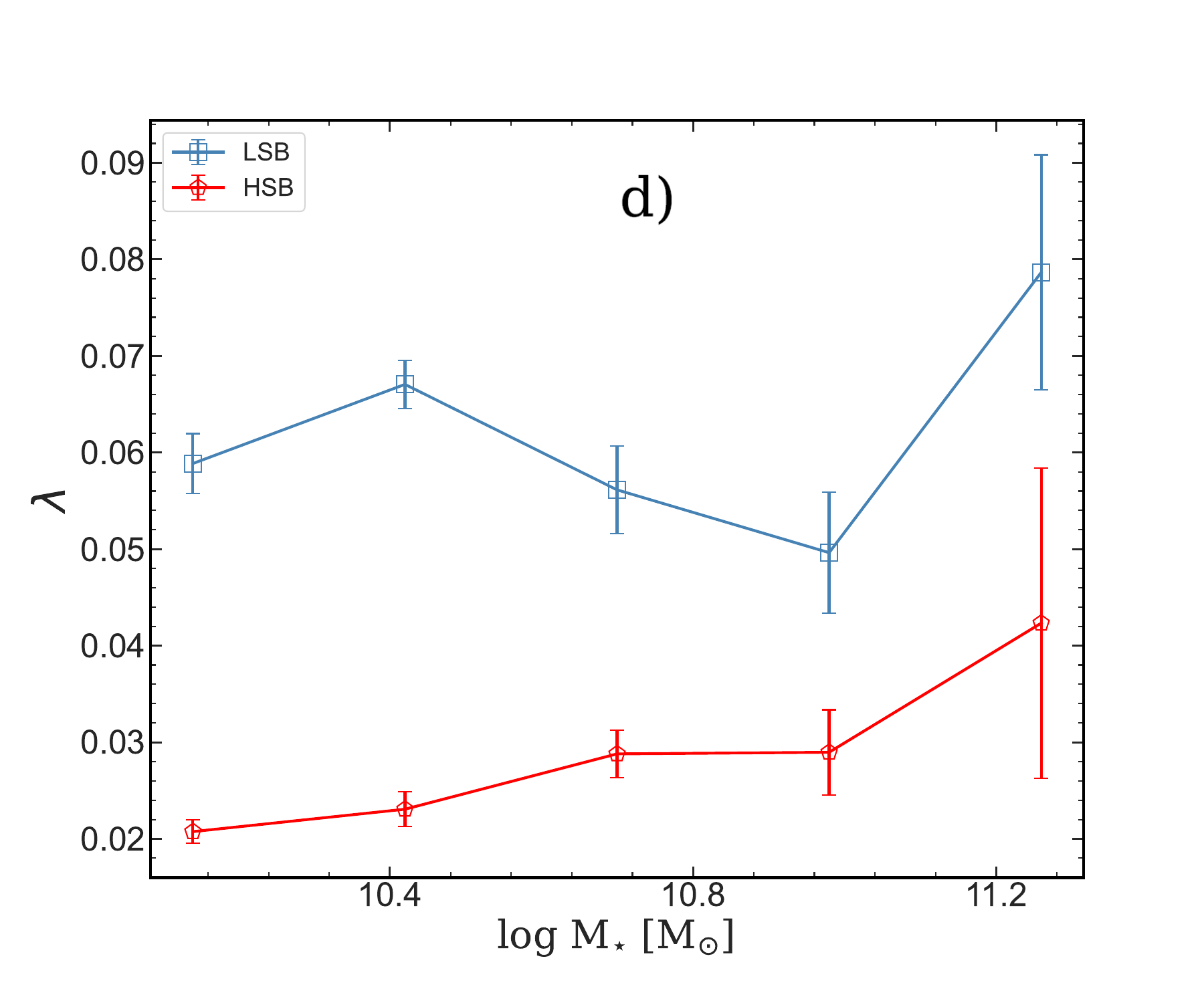}
\vspace{-0.6cm}
\caption{ }
    \label{subflambdaMs}
\end{subfigure}

\vspace{-0.12cm}

\begin{subfigure}{0.45\linewidth}
    \includegraphics[width=\linewidth, height=0.85\linewidth]{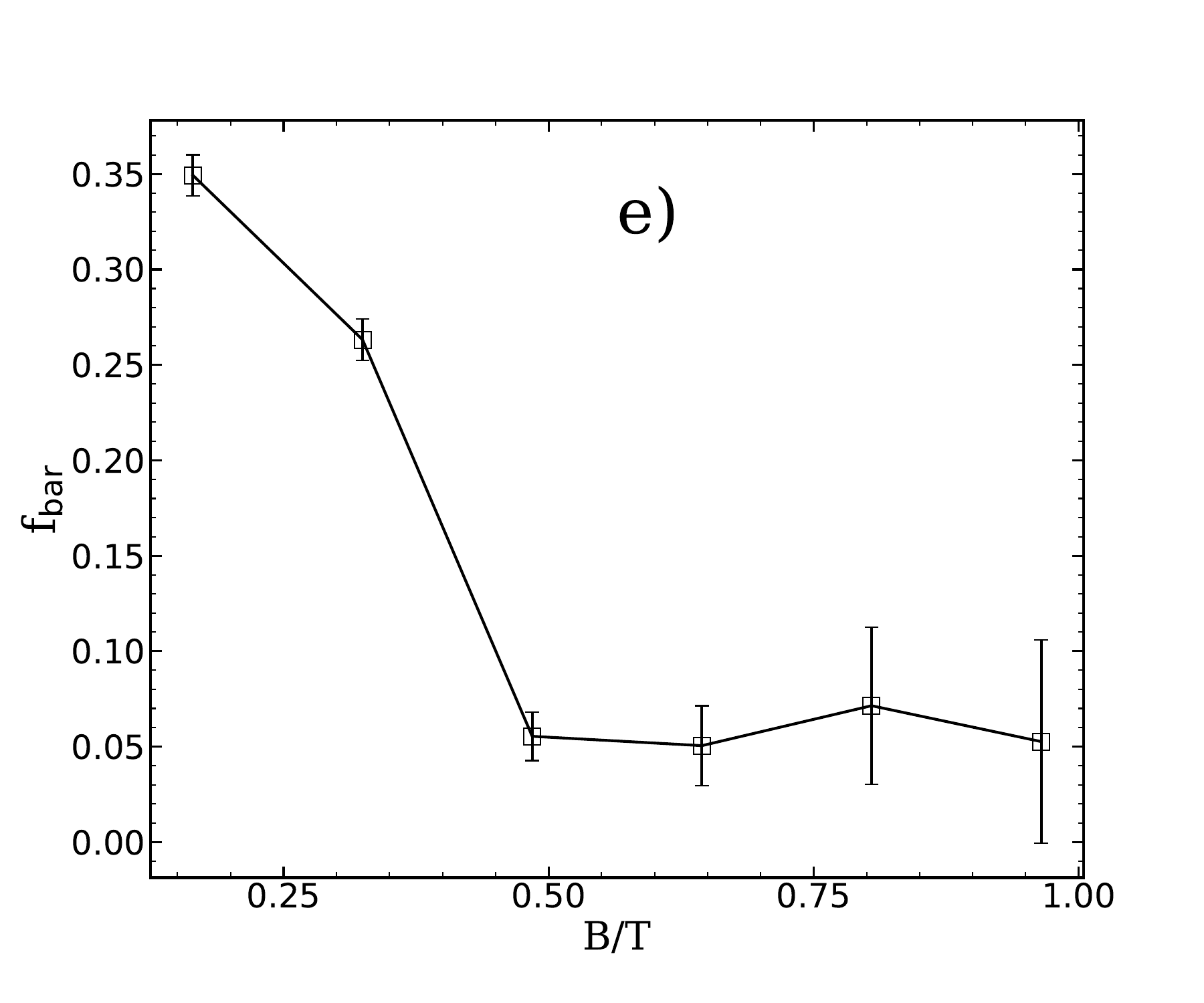} 
\vspace{-0.6cm}
\caption{ }
    \label{totalfbarBT}
\end{subfigure}
    \hspace{.8cm}
\begin{subfigure}{0.45\linewidth}
    \includegraphics[width=\linewidth, height=0.85\linewidth]{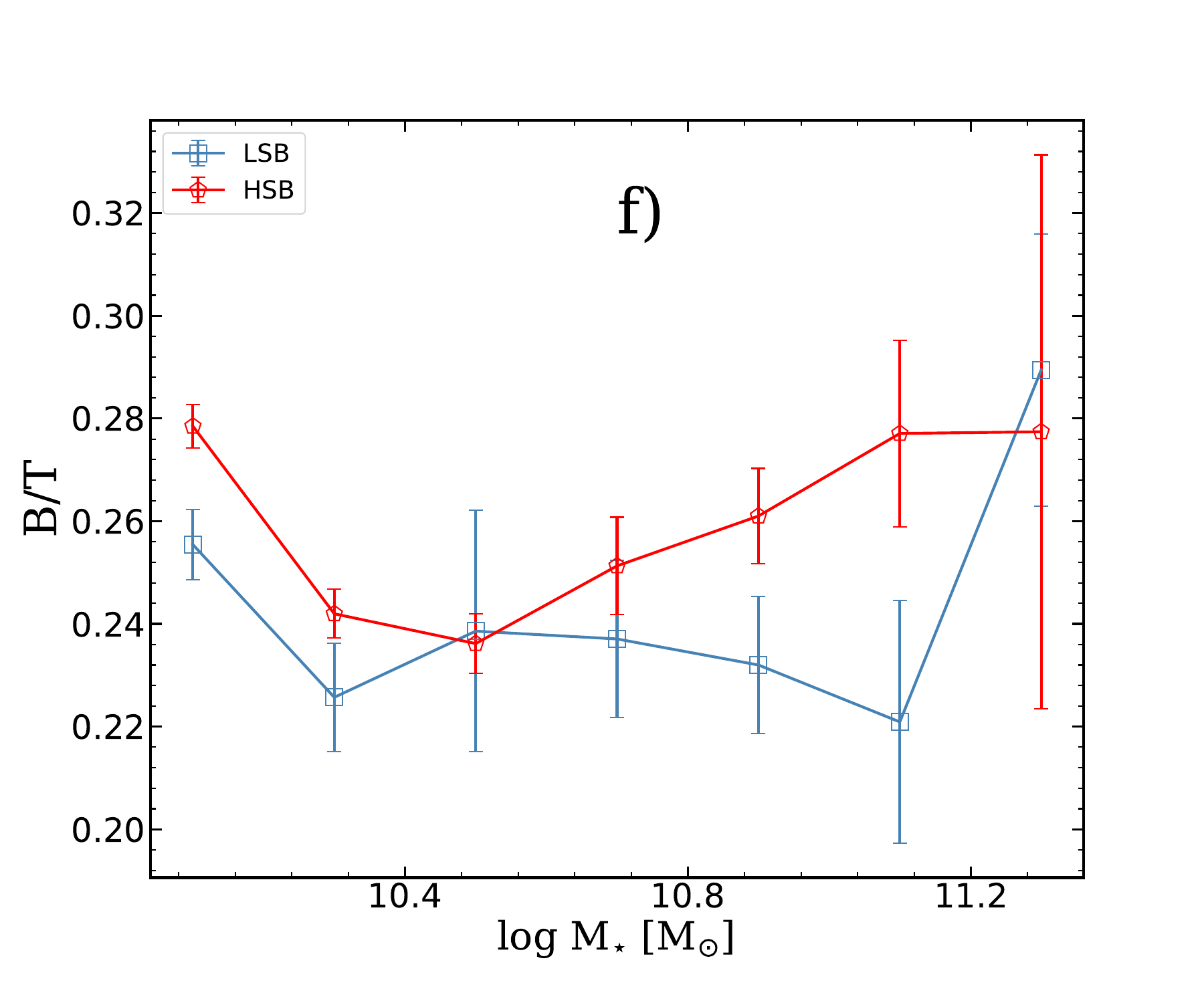}
\vspace{-0.6cm}
\caption{ }
    \label{subBTMs}
\end{subfigure}

\caption{Left panels depict the bar fraction as a function of a) gas fraction, c) spin parameter, and e) bulge-to-total mass, considering the parent galaxy sample. Right panels present the behavior of the same parameters as a function of stellar mass for both LSBs and HSBs samples. The plots in the left panels reveal that the bar fraction tends to decrease as both $f_{\mathrm{gas}}$ and $B/T$ increase. Regarding the spin parameter, the most significant change is observed in the low stellar mass range, where a notable decrease in $f_{\mathrm{bar}}$ occurs with an increase in $\lambda$. The right panels indicate that LSBs exhibit higher gas fractions and larger spin parameters compared to HSBs, also featuring less prominent bulges.}
\label{fbar-par}%
\end{figure*}

In this section, we explore the behavior of the bar fraction with the different galaxy parameters beyond the stellar mass, which is also related to the presence of a bar. The behavior of bar fraction as a function of these parameters (gas fraction,  spin parameter, and bulge-to-total mass) is presented in  Fig. \ref{fbar-par} for the total sample, as well as for LSBs and HSBs. In the panel \ref{totalfbarfgas}, we present $f_{\rm bar}$ as a function of the gas fraction. Here, we can see that the bar fraction is strongly dependent on the gas content since $f_{\rm bar}$ shows a negative trend when increasing the gas content in a galaxy. The decrease in $f_{\rm bar}$ is particularly pronounced for small values of $f_{\rm gas}$ (up to $\sim 0.25$). On the right top panel (\ref{subfgasMs}), we present the gas fraction as a function of stellar mass, where it is appreciated that on average, LSBs have a $f_{\mathrm{gas}}$ three times higher than HSBs. For both HSB and LSB samples, we also see that the gas fraction decreases towards higher stellar masses. 

Panel \ref{totalfbarlambda} shows the bar fraction as a function of the spin parameter $\lambda$. It is observed that $f_{\mathrm{bar}}$ decreases as the spin parameter increases in the range $0 < \lambda < 0.07$, falling from a value higher than 30\% to close to $\sim 20\%$, stabilizing at this value for $\lambda \geq 0.08$. For higher values than $\sim 0.08$, no significant change is noticeable. This indicates that, at least for our sample, there is a negative dependence of $f_{\mathrm{bar}}$ with the spin value mainly on the range of small $\lambda$. 

On the other hand, in Fig. \ref{subflambdaMs}, we show $\lambda$ as a function of the stellar mass for subsamples. As can be seen, LSBs have systematically higher values of $\lambda$ throughout all the mass ranges, compared to the values of HSBs as reported by \cite{perez2022formation, perez2024environmental}. Although an increase in the spin parameter of HSBs is observed from $\sim 10^{10.4}$ M$_{\odot}$, they do not exceed the values of their low-brightness counterparts. Given that high spin values suppress the presence of stellar bars, it is natural to expect a lower fraction of barred galaxies for LSBs compared to HSBs when considering only the spin parameter. It is important to mention that the results shown consider the total spin; however, we obtain the same results when we consider the counterpart of the DM-only simulation.

Finally, in the bottom panels, we found $f_{\mathrm{bar}}$ as a function of bulge-to-total mass fraction (Fig. \ref{totalfbarBT}) and $B/T$ as a function of stellar mass (Fig. \ref{subBTMs}). As we can see, in our total sample, there are galaxies with high values of $B/T$ exceeding 0.6, some even approaching 1. This appears to be inconsistent with the sample description, which considers disc-dominated galaxies where no prominent bulge components are expected. However, it is important to note that the bulge values were obtained from a different source and classification than the $\kappa_{\rm rot}$ parameter, which we used to define a galaxy as disc-dominated. Furthermore, the number of galaxies with $B/T \geq 0.6$ is very small, accounting for only 1.5\% and 3.2\% of LSBs and HSBs, respectively. Additionally, these galaxies with high $B/T$ values also have $\kappa_{\rm rot}$ values less than 0.6. This indicates that there are very few galaxies in the total sample that exhibit a discrepancy in their morphological classification.

From Fig. \ref{totalfbarBT}, a notable decrease in $f_{\mathrm{bar}}$ is observed with increasing $B/T$ ratio, particularly at low values of $B/T< 0.5$. Beyond this decrease, the bar fraction levels off and remains relatively constant for higher bulge-to-total fraction values. 
In Fig. \ref{subBTMs}, we observe a gradual increase in $B/T$ for HSBs with stellar masses higher than $\sim 10^{10.5}$ M$_{\odot}$ reaching its highest point at $B/T \sim 0.3$ for the most massive galaxies. However, this is accompanied by larger uncertainties due to a statistically low number of galaxies. In contrast, LSBs exhibit a decreasing trend, with a drop in the bulge fraction up to a stellar mass of $10^{11}$ M$_{\odot}$. 

\subsection{The bar fraction in control samples}
\label{barfrac-CS}
\begin{figure*}
\captionsetup[subfigure]{labelformat=empty}
\begin{subfigure}{0.45\linewidth}
    \includegraphics[width=\linewidth, height=0.8\linewidth]{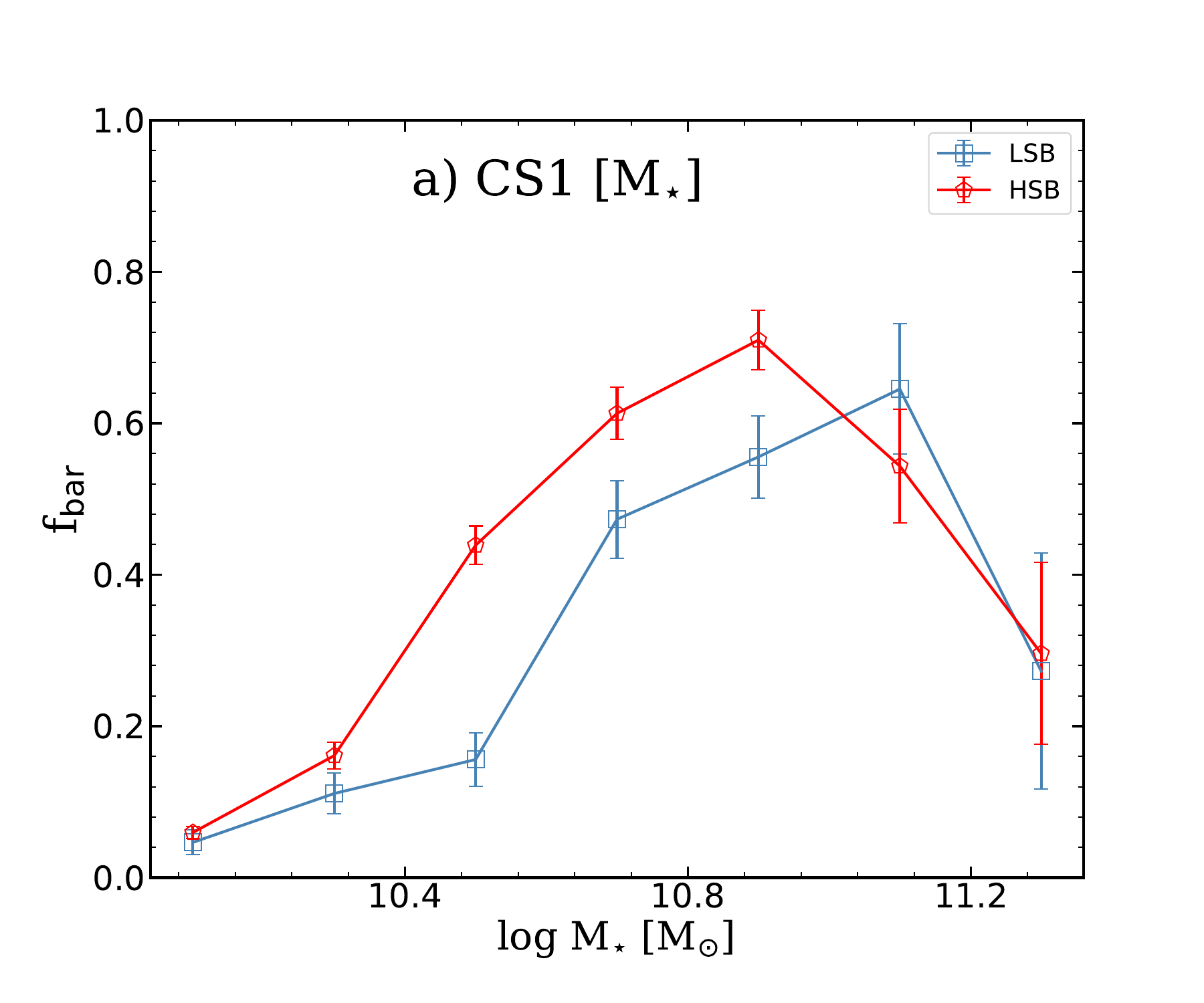}
\vspace{-0.4cm}
\caption{  }
    \label{fig:a} 
\end{subfigure}
    \hspace{.8cm}
\begin{subfigure}{0.45\linewidth}
    \includegraphics[width=\linewidth, height=0.8\linewidth]{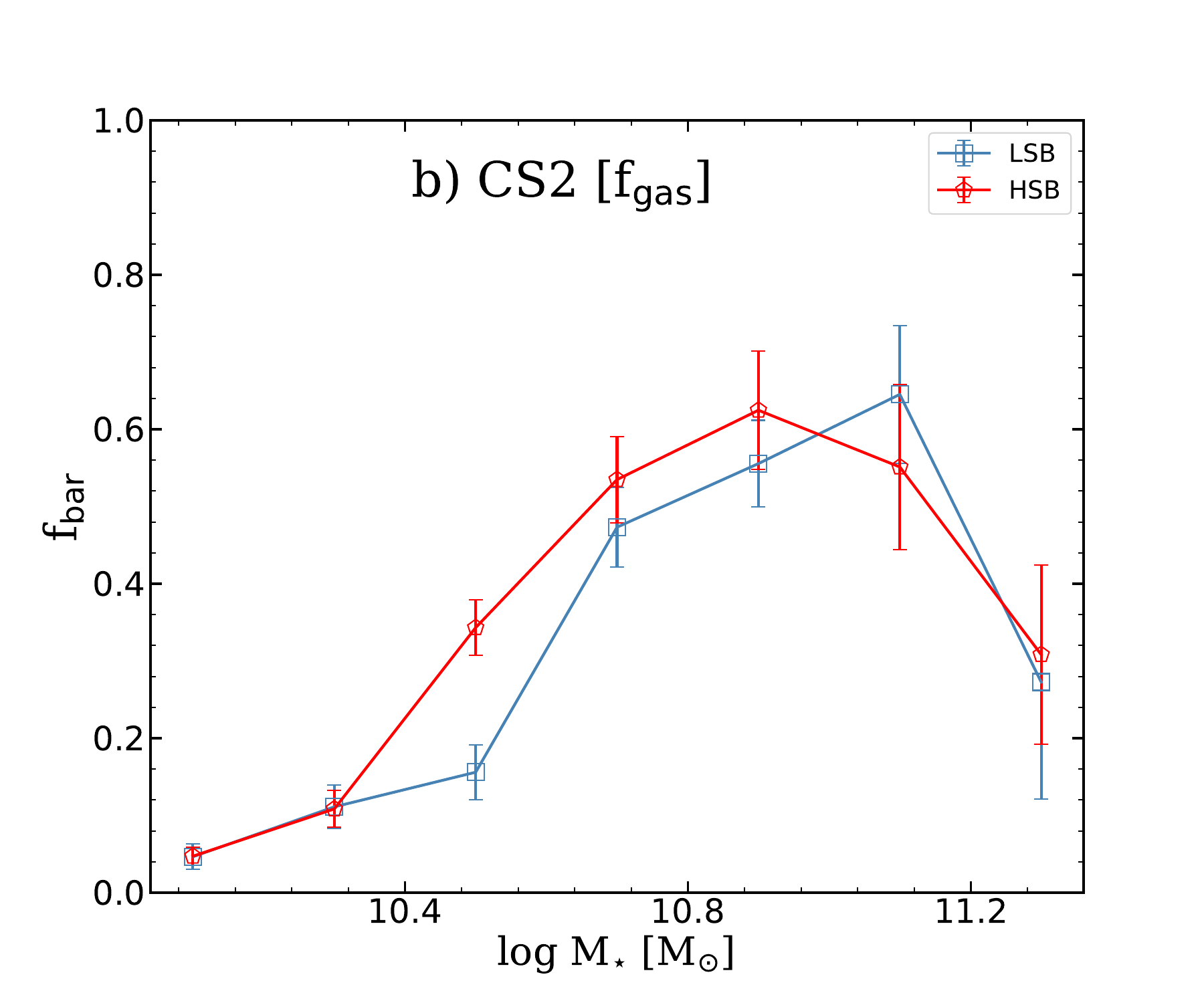}
\vspace{-0.4cm}
\caption{  }
    \label{fig:b}
\end{subfigure}

\begin{subfigure}{0.45\linewidth}
    \includegraphics[width=\linewidth, height=0.8\linewidth]{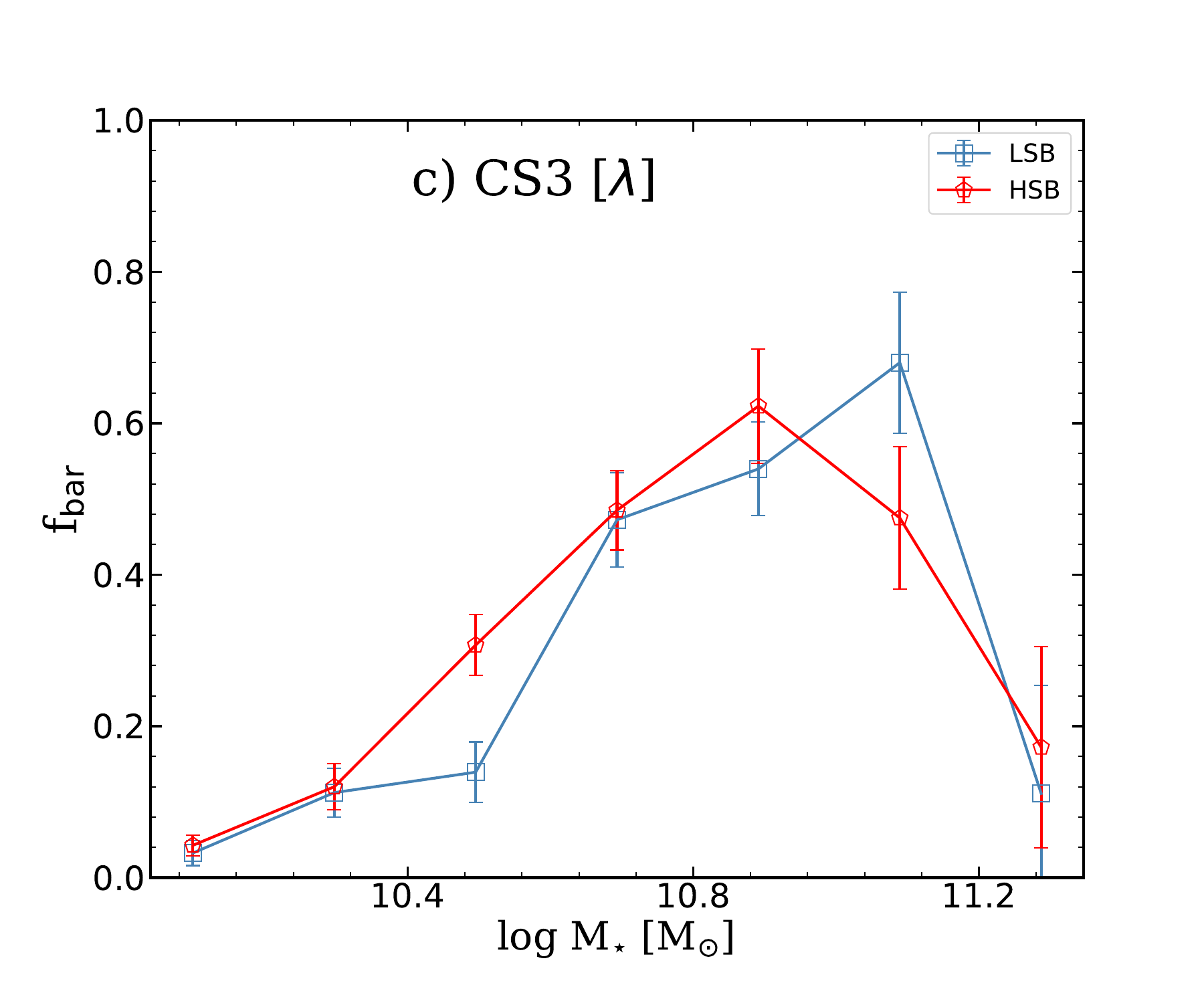}
\vspace{-0.4cm}
\caption{  }
    \label{fig:c}
\end{subfigure}
    \hspace{.8cm}
\begin{subfigure}{0.45\linewidth}
    \includegraphics[width=\linewidth, height=0.8\linewidth]{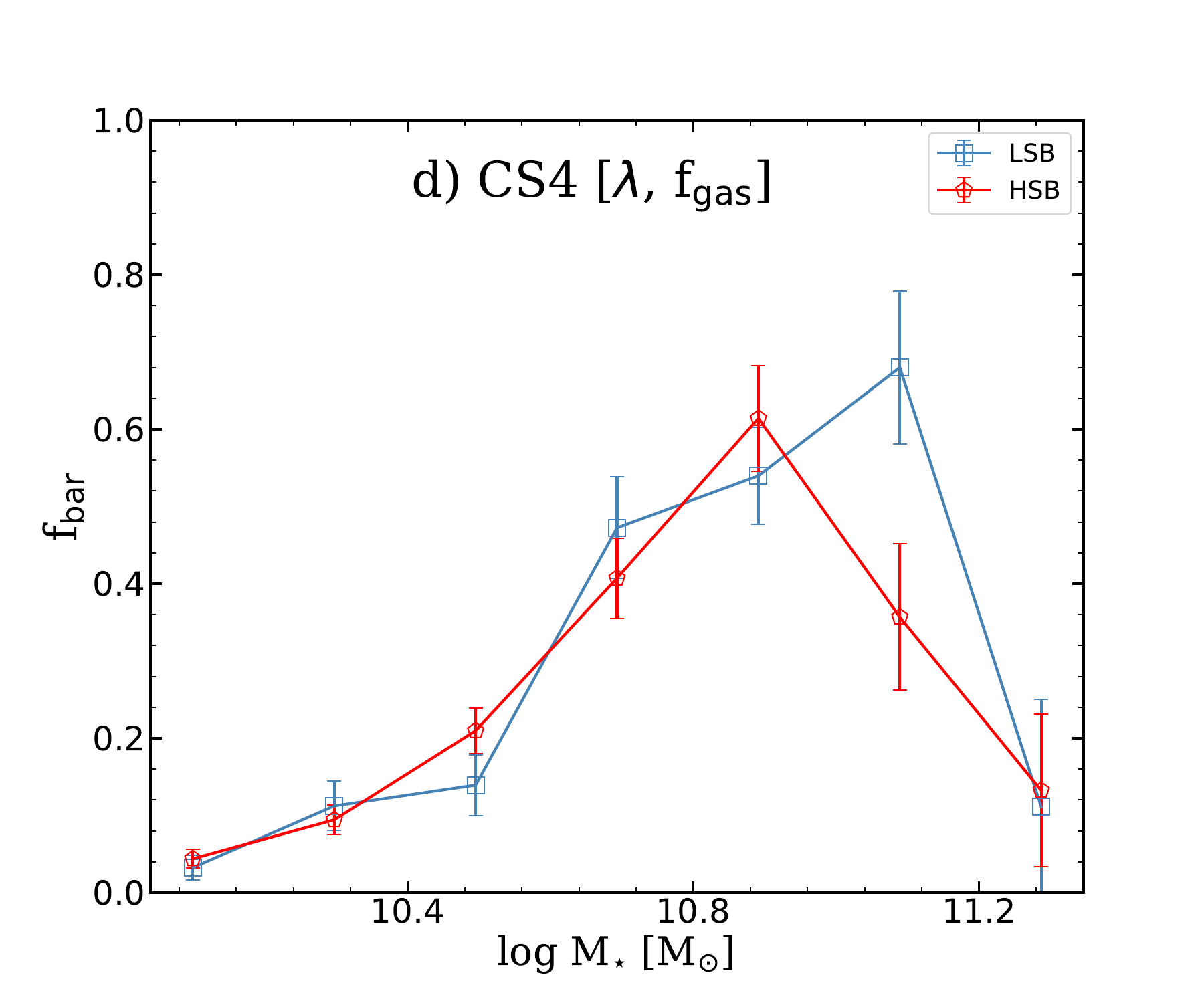} 
\vspace{-0.4cm}
\caption{  }
    \label{fig:d}
\end{subfigure}

\begin{subfigure}{0.45\linewidth}
    \includegraphics[width=\linewidth, height=0.8\linewidth]{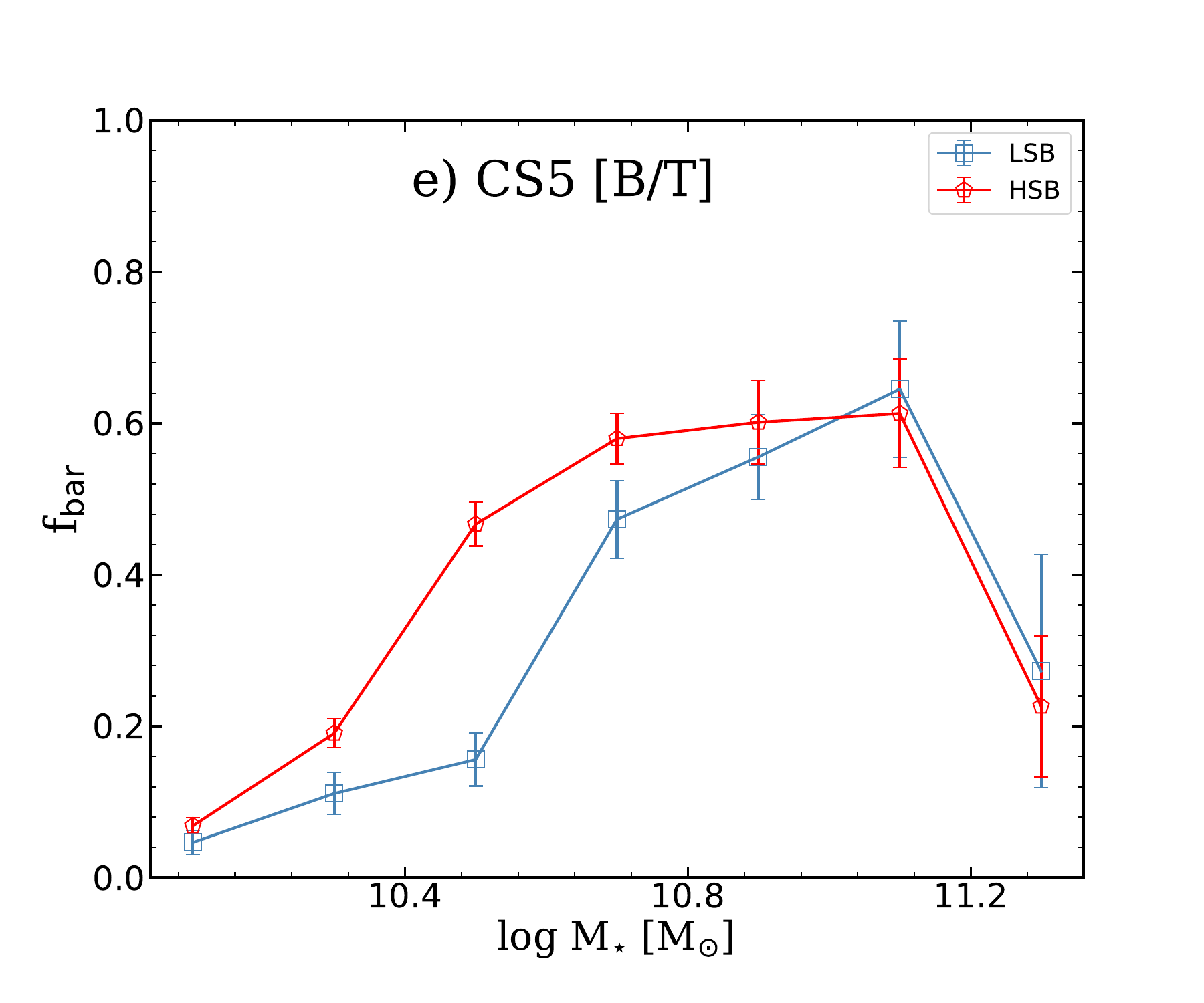}
\vspace{-0.4cm}
\caption{  }
    \label{fig:e}
\end{subfigure}
    \hspace{.8cm}
\begin{subfigure}{0.45\linewidth}
    \includegraphics[width=\linewidth, height=0.8\linewidth]{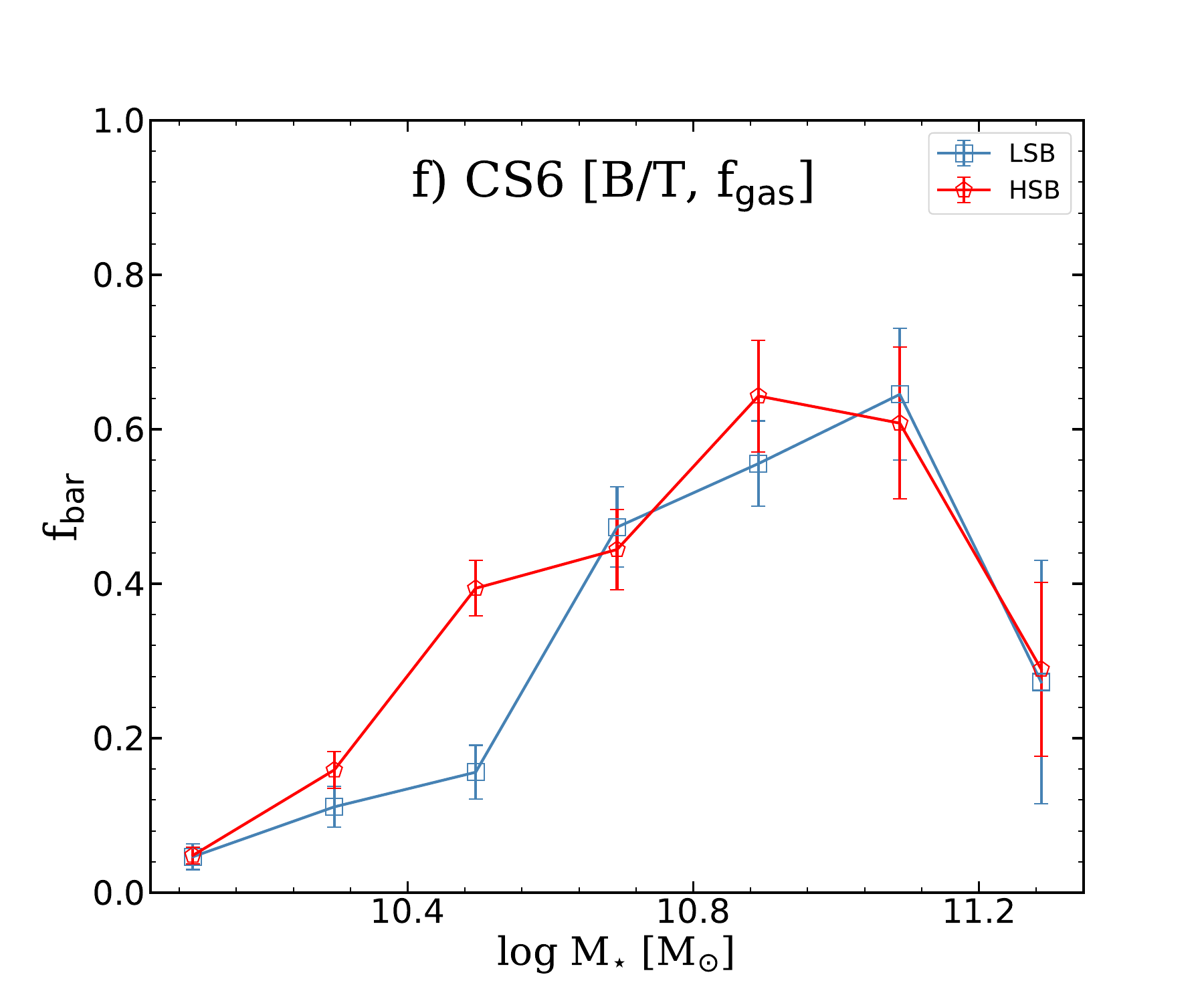} 
\vspace{-0.4cm}
\caption{  }
    \label{fig:f}
\end{subfigure}

\caption{Bar fraction as a function of stellar mass for six control samples, considering stellar mass, gas mass fraction, bulge-to-total mass fraction, and the spin. It becomes apparent that control samples demonstrating the most convergence in bar fraction between LSBs and HSBs are those involving the spin parameter and gas mass fraction, at low and intermediate stellar mass ranges, particularly in CS4 (panel d). Additionally, the control sample that best converges the bar fraction across the high mass range is associated with bulge-to-total mass fraction, as evidenced in CS6 (panel f).} 
\label{CS}
\end{figure*}

In this section, we present the results of applying the various control samples defined in section \ref{CS-desc}. The changes in the bar fraction for the first control samples (CS1 to CS6), which considers only physical parameters, are illustrated in Figure \ref{CS}.
 
As observed in Fig. \ref{fig:a} for CS1, where only stellar mass is controlled, the bar fraction for HSBs is higher than that for LSBs in the low and intermediate-mass ranges. However, for M${\star} \geq 10^{11}$ M${\odot}$, there is no significant difference in the bar fraction between the LSB and HSB subsamples. This suggests that, at least for our sample in this mass range, the stellar mass of the galaxy is not the primary parameter responsible for the difference in the bar fraction of LSBs and HSBs.

For Figs. \ref{fig:b} \& \ref{fig:c},  corresponding to CS2 and CS3, where the controlled parameters are the gas mass fraction and the spin parameter respectively, we observe a decreasing difference in bar fraction for galaxies with M$_{\star} < 10^{11}$ M${\odot}$. Although the distinction is less pronounced, the bar fraction in LSBs is still lower than in HSBs in galaxies with stellar masses around M$_{\star} \sim 10^{10.5}$ M$_{\odot}$. In the high mass range, for CS2 there is no significant difference in the bar fraction between LSB and HSB, compared to the uncontrolled sample, but for CS4, a difference between the bar fractions emerges, LSB surpassing HSB. This behavior indicates that both the gas content and the spin parameter individually have a significant effect on the bar fraction differences between LSBs and HSBs for low and intermediate stellar mass ranges. When we combine the effects of the gas mass fraction and the spin parameter (Fig. \ref{fig:d}), we can see that there is no difference in $f_{\mathrm{bar}}$ for subsamples in the low and intermediate-mass ranges. However, in the high mass range, we observe a substantial discrepancy in the bar fraction for the subsamples. The combination of these parameters suggests that the difference in the prevalence of bars in the subsamples is strongly influenced by these physical parameters for galaxies with stellar mass smaller than $10^{11}$ M$_{\odot}$. However, the discrepancy in the high mass range indicates that there is another factor that could affect the bar fraction, such as the bulge, which is considered in the following control samples.

The control sample CS5 (Fig. \ref{fig:e}) considers the bulge mass fraction, exhibiting an impact on the bar fraction on the opposite mass range when compared to CS3 and CS4. Within CS5, we observe a divergence in bar fraction between LSBs and HSBs for stellar mass ranges below $10^{11}$ M$_{\odot}$, mirroring the observations in Fig. \ref{fbar-logM}, where in HSBs exhibit a higher bar fraction than LSBs. Beyond this threshold, a noticeable convergence in bar fractions is evident, suggesting that, in the high mass range, the bulge component significantly influences the presence of bars. However, for the low and intermediate-mass range, the bulge does not exert a substantial impact on $f_{\mathrm{bar}}$.

Finally, in CS6, as illustrated in Fig. \ref{fig:f}, where both the bulge mass fraction and gas mass fraction are controlled, we obtain a compromise for the convergence in bar fractions for our subsamples, in which the parameter that mostly affects galaxies in the intermediate and low mass end is the gas content, whereas, for the stellar mass end, the bulge prominence seems to be the most important parameter affecting the bar fraction. For lower stellar masses, a discernible difference in bar fractions between LSBs and HSBs persists, albeit less prominently compared to the trends depicted in Fig. \ref{fbar-logM} for the same stellar masses. 

\subsection{Dependence of bar fraction with galaxy environment}

\begin{figure*}
\begin{subfigure}{0.49\linewidth}
    \includegraphics[width=\linewidth, height=0.8\linewidth]{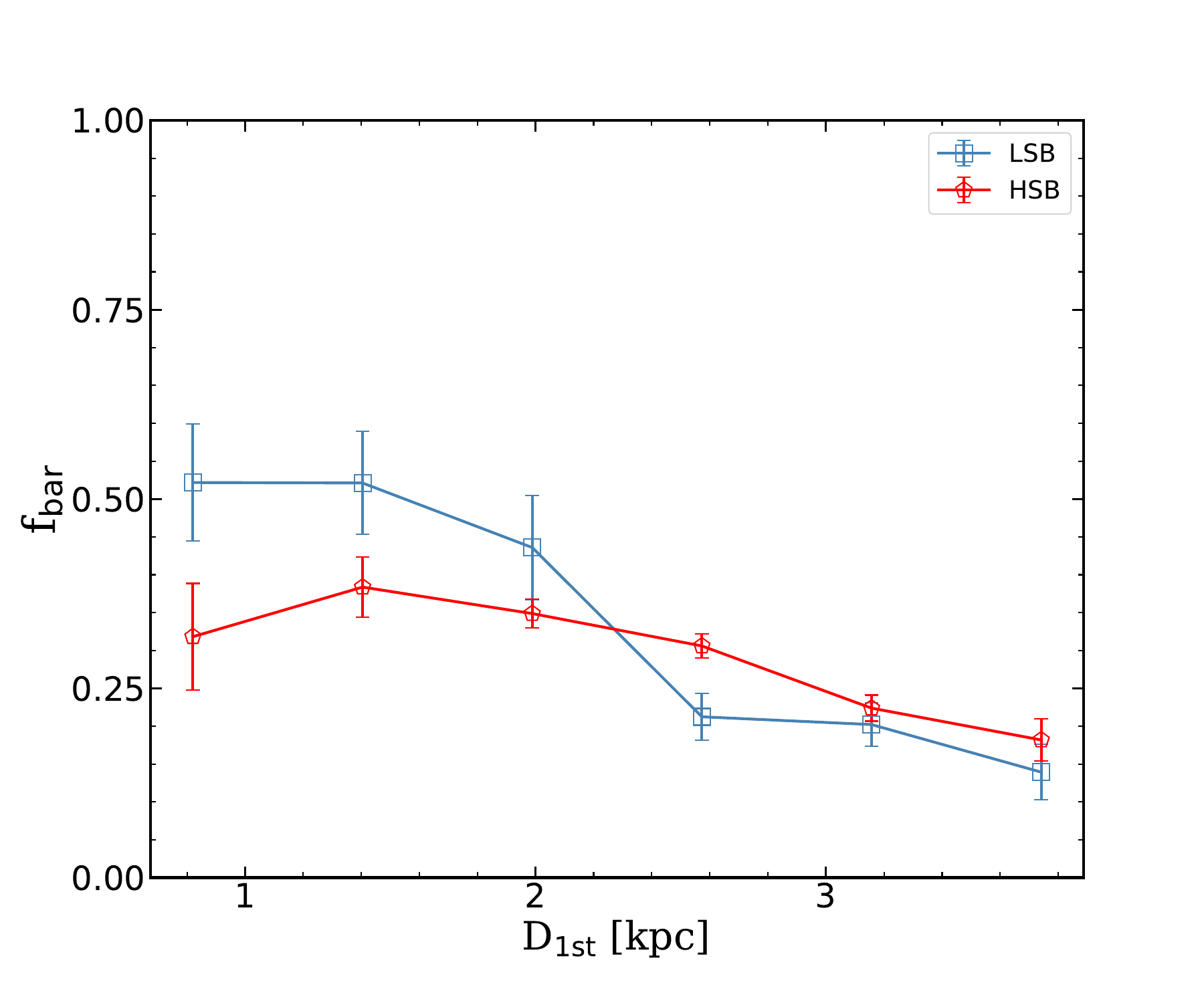}
    \label{fig:c1}
\end{subfigure}
    \hfill
\begin{subfigure}{0.49\linewidth}
    \includegraphics[width=\linewidth, height=0.8\linewidth]{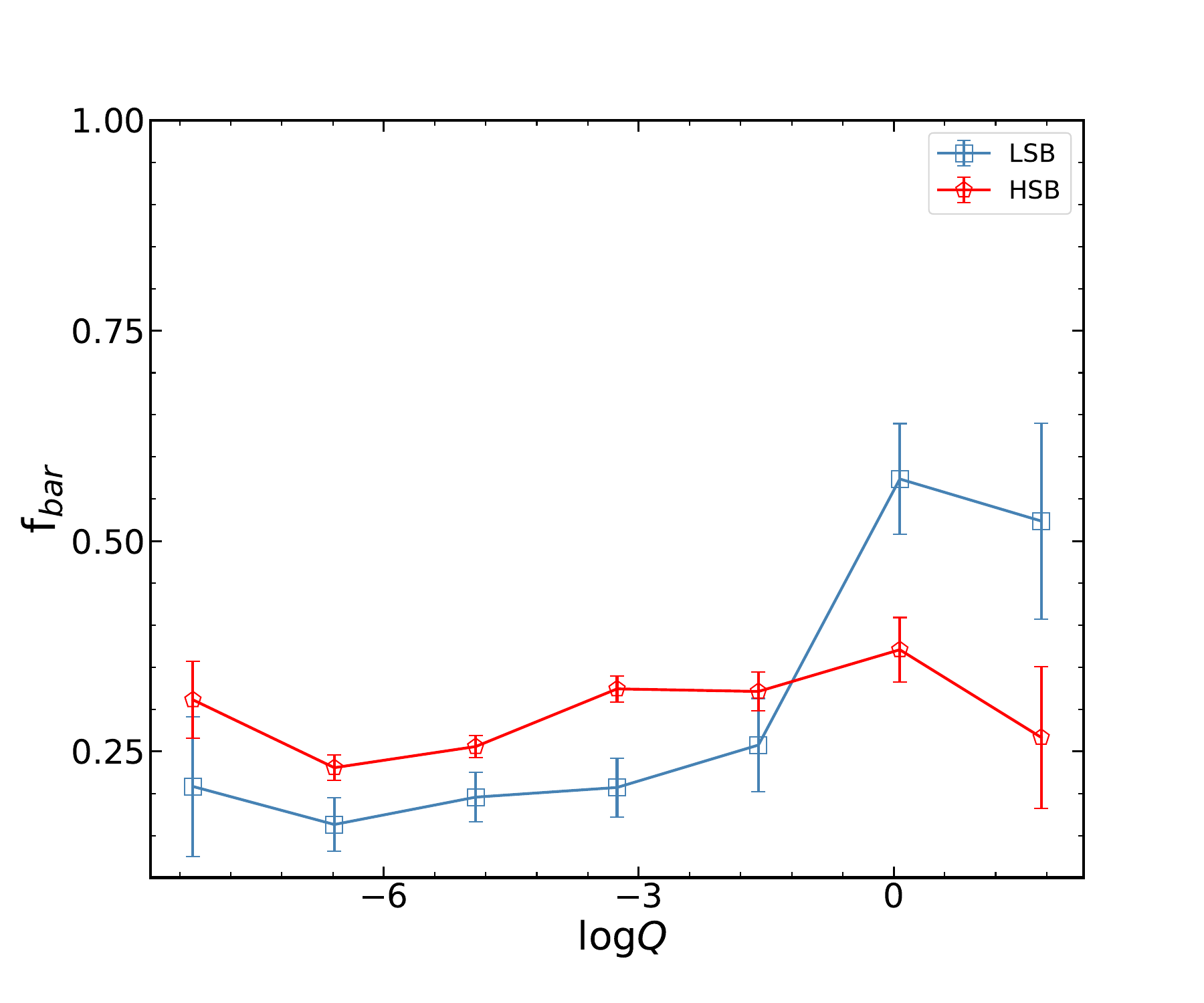}
    \label{fig:d1}
\end{subfigure}

\caption{Bar fraction as a function of distance to the nearest neighbor (left panel), and tidal parameter (right panel). A distinct decrease in the bar fraction is observed with increasing distance to the nearest neighbor, particularly notable in the case of LSBs. A more pronounced increase is evident when examining the tidal parameter, showcasing a dramatic rise in the bar fraction for LSBs, especially in strongly interacting systems.} 
\label{environmentplots}
\end{figure*}

In this section, we explore the relationship between the bar fraction and the local environment. We use the tidal parameter (Eq. \ref{eqQ})  to measure the interaction strength. We also use the distance to the nearest neighbor, D$_1$, to assess the environmental impact.  We observe in Fig. \ref{environmentplots} that the bar fraction decreases as the distance to the closest galaxy increases. For LSBs, this drop is especially pronounced in the intermediate distance range, while for HSBs, although a negative trend with distance exists, it is less prominent.  

In the right panel, we present the dependence of the bar fraction on the tidal parameter $Q$. HSBs exhibit a relatively constant bar fraction across the entire range of $Q$, while LSBs show a dramatic increase of $f_{\mathrm{bar}}$ for high $Q$ values. The bar fraction notably increases in LSBs with higher tidal parameters ($\geq -2$), indicating greater sensitivity to environmental influences. Our results suggest that, in LSBs, environmental factors, such as strong tidal interactions, play an important role in the presence of stellar bars.

In summary, LSBs appear to be significantly more affected by neighbors and tidal interactions as expected (\citealt{moore1999fate, das2013giant, di2019nihao}), leading to an increased bar fraction with stronger interactions. This susceptibility may be due to instabilities in their sparse, lower-density discs. Conversely, HSBs, while also influenced by environmental factors, exhibit a less significant dependence on the bar fraction.

\subsubsection{Bar fraction for the tidal parameter control sample.}

\begin{figure}
\includegraphics[width=\columnwidth]{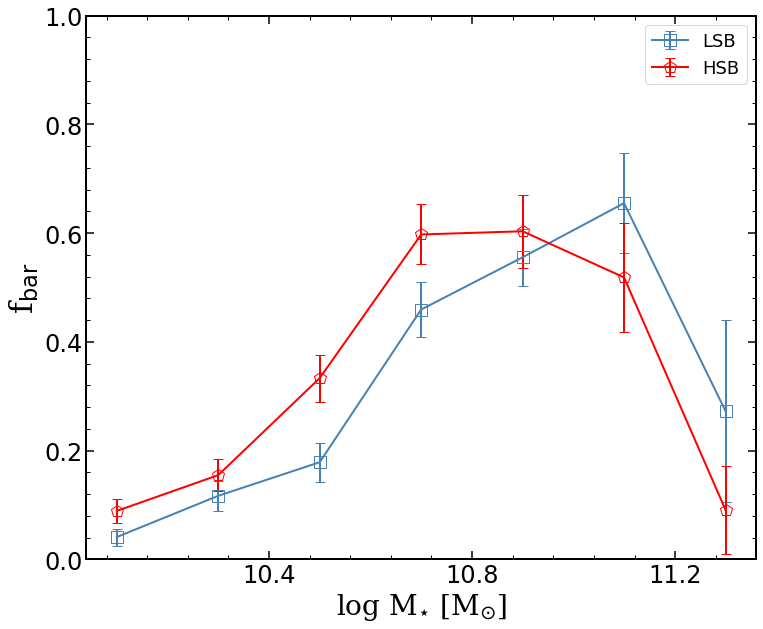}
\caption{The bar fraction as a function of stellar mass for the control sample (CS7) where the tidal parameter Q is considered. As observed, this external parameter seems to have no discernible impact on the difference of the bar fraction between LSBs and HSBs.} 
\label{CS-Q}
\end{figure}

To investigate the impact of the environment on the bar fraction, we established a final control sample (CS7) employing only the tidal parameter $Q$ as the controlling parameter. The results are depicted in Fig. \ref{CS-Q}, revealing that for low and intermediate stellar masses (down to $10^{11}$ M${\odot}$), the bar fraction exhibits a noticeable distinction between LSBs and HSBs, with higher bar fraction values for HSBs compared to LSBs. However, this difference, though appreciable, is smaller than the contrast seen in our original uncontrolled sample presented in Fig. \ref{fbar-logM}. At the high-mass end (M${\star} \sim 10^{11}$ M$_{\odot}$), LSBs display higher bar fractions than HSBs, though considering the confidence intervals, these differences are less significant, especially for more massive galaxies. This pattern suggests that tidal interactions have a limited impact on the bar fraction. When controlling for it, the difference in bar fractions between LSBs and HSBs at fixed stellar mass is reduced.

\section{Discussion}
\label{discussion}

In this work, we employ the cosmological magneto-hydrodynamical simulation TNG100 from the IllustrisTNG project to investigate and analyze the presence of stellar bars in LSBs and HSBs. Our analysis is based on a sample of $4,224$ disc galaxies with stellar mass M${_\star} \geq 10^{10}$ M${\odot}$ at $z=0$. To identify bar structures in our galaxies we employ the catalogue by \cite{zhao2020barred},  based on the ellipse fitting method. To segregate between LSBs and HSBs, we use the selection criteria employed by \cite{perez2022formation}, which is based on a limit value for central surface brightness of $22$ mag arcsec$^{-2}$ in the $r$ band. With this condition, we identify $3, 572$ HSBs and $652$ LSBs. The low fraction of LSBs in our sample is expected, given that LSBs are more abundant in the low-mass regime (\citealt{dalcanton1997number}, \citealt{galaz2011low} and \citealt{perez2022formation}). The main results obtained are presented below. 

Applying the criteria for bar identification, we identify  $1, 172$ barred galaxies that represent a bar fraction of $27.75 \pm 0.68 \%$, a result consistent with observational (\citealt{masters2011galaxy}, \citealt{lee2012dependence} and \citealt{vera2016effect}) and theoretical studies that make use of cosmological simulations (\citealt{algorry2017barred}, \citealt{peschken2019tidally}, \citealt{rosas2022evolution}). For the LSB and HSB subsamples, the fraction of barred galaxies corresponds to $24.73$ $\pm 1.73 \%$ and $28.35 \pm 0.74 \%$, in line with the results reported by \cite{mcgaugh1995morphology}, \cite{impey1996low} and \cite{sodi2017low} where bars appear to be less abundant in LSBs. The difference of the bar fraction remains even when the sample is controlled by their stellar mass, being systematically higher in HSBs, except for galaxies with a stellar mass exceeding $\sim 10^{11}$ M$_{\odot}$, where the difference vanishes (see Fig. \ref{fig:a}). \\

To identify the specific parameters influencing the bar fraction within our sample across the considered range of stellar masses, we conducted an analysis of both physical (stellar mass M$_{\star}$, gas mass fraction f$_{\rm gas}$, bulge-to-total fraction $B/T$ and adimensional spin $\lambda$), and external  (tidal parameter $Q$ between the galaxy and its nearest neighbor) factors. The selection of these parameters is motivated by numerous studies indicating their potential influence on the bar fraction and mentioned in the introduction and in section \ref{sect-fbar-par}.

\subsection{Bar fraction as a function of physical parameters.}

Regarding the bar fraction, several observational works (\citealt{oh2011bar, diaz2016characterization, sodi2017stellar}) and theoretical studies (\citealt{rosas2020buildup}, \citealt{zhao2020barred}) have reported its increase with increasing stellar mass. These theoretical studies, employing the IllustrisTNG cosmological simulation, show that the highest value for bar fraction occurs at M$_{\star} \sim 10^{10.9}$ M$_{\odot}$. Above this value, the fraction slowly decreases to the high mass end. A compilation of this behavior for different studies is present in \cite{roshan2021fast}. For our sample, the behavior of the bar fraction with stellar mass aligns with the literature as illustrated in Fig. \ref{fbar-logM}, where the sample exhibits a peak bar fraction of approximately $ \sim 60 \%$ at a stellar mass of $10^{10.9}$ M$_{\odot}$, followed by a gradual decline for larger masses. \\

Concerning the dependence of the bar fraction on the galaxies gas content, numerous observational studies \citep{masters2012galaxy,sodi2017stellar, zhou2021correlation}  have demonstrated that the presence of gas could inhibit the formation and growth of bars. In gas-rich galaxies, a significant angular momentum exchange occurs between the gas and stars (\citealt{combes1993bars, athanassoula2003determines}). This exchange increases the bar rotation frequency (\citealt{villa2010dark}), subsequently weakening and ultimately disrupting it. \cite{athanassoula2013bar} also concluded that in simulated gas-rich systems, bar growth is slower compared to models with lower gas content. We found similar behavior in our sample, with the bar fraction declining from $50 \%$ to $\sim 0\%$ toward higher gas fractions. However, other studies analyzing high-resolution zoom-in simulations show that for high redshift objects, gas under certain conditions and with frequent perturbations (such as mergers and close flybys) favors the bar formation (\citealt{bi2022emergence, bi2022modeling}). 

Our third physical parameter is the dimensionless spin parameter. Both theoretical (\citealt{saha2013spinning}, \citealt{long2014secular}) and observational (\citealt{cervantes2013galactic}) studies have shown that an increase in spin leads to the quenching of both the size and strength of bars, as a high spin value hinders effective angular momentum transfer, thereby impeding bar growth. This pattern is also evident in our parent sample, as depicted in Fig. \ref{totalfbarlambda}, where the bar fraction decreases for higher spin values. The decline is more pronounced for lower values ($\lambda< 0.06$), where the bar fraction transitions from approximately $\sim 30 \%$ to $\sim 20 \%$. According to \cite{li2023stellar}, which uses numerical simulations, for high spin values the buckling instability is postponed, and after this phenomenon, the bar regains strength progressively slower. Additionally, \cite{li2024evolution} found that a faster-rotating halo decreases the amplitude of the bar.  Furthermore, \cite{collier2019stellar} found that increasing counter-rotating $\lambda$ decreases the bar instability, which aligns with the findings of \cite{saha2013spinning}. Their study also concluded that the impact of spin depends on the orientation of the halo. Specifically, an increase in retrograde $\lambda$ has a lesser effect on bar evolution compared to an increase in prograde $\lambda$, where bars tend to be dampened (\citealt{collier2018makes}). \cite{li2023stellar}, using numerical simulations, concluded that the density of the rotating dark matter halo in which the galaxy is embedded plays a significant role in bar evolution. For high spin values and low densities, the buckling instability of stellar bars is postponed, with the effect being more pronounced as the density decreases for fixed values of $\lambda$.
 
Finally, we investigate the impact of the bulge component on the bar fraction. This relationship has been explored in theoretical studies, such as \cite{weinzirl2009bulge}, which observed a decrease in the bar fraction from $68 \%$ to approximately $36 \%$ when $B/T > 0.2$. This suggests that low bulge-to-total ratios favor bar instabilities, making bars more prevalent under these conditions. A similar trend was confirmed by \cite{kataria2018study}, who, using N-body simulations, concluded that massive and more concentrated bulges can inhibit bar formation and growth. Specifically, no bar formation occurs for bulge-to-disc mass fractions greater than $0.6$. Using zoom-in simulations of barred galaxies in TNG50 simulations, \cite{rosas2024galaxy} found similar limits, even though the disc could be bar unstable. This result aligns with the findings of \cite{saha2018some}, who demonstrated that compact classical bulges prevent bar formation in cold stellar discs in collisionless simulations. In an observational study by \citealt{yoon2019observational}, the authors noted a decline in the bar fraction with increasing $B/T$ when analyzing barred galaxies in the cluster environment. This reduction in the bar fraction as the bulge component increases is also evident in our sample, as shown in Fig. \ref{totalfbarBT}. Examining the evolution of bars, \cite{rosas2022evolution} found that the bulge component in barred galaxies is generally small and already present in the host galaxy before bar formation. Once the bar forms, the bulge does not undergo significant growth. Consequently, discs with small bulges are more prone to forming bars compared to discs with substantial bulges. Similarly, \cite{izquierdo2022disc}, in the IllustrisTNG simulation, observed that the bar could contribute to the growth of the bulge.

\subsection{Bar fraction as a function of local environment.}

In addition to the physical parameters mentioned above, we investigated the impact of the tidal parameter $Q$ on the bar fraction (Fig. \ref{environmentplots}). For HSBs, the bar fraction appears independent of the tidal parameter. However, for LSBs, we observe a gradual increase at low values of the tidal parameter. In the range of $\log Q \geq -2$, the bar fraction surges to approximately $0.55$, indicating that strong interactions can substantially boost the bar prevalence in these galaxies. Given the extended nature and lower stellar density of LSBs, they are more susceptible to strong tidal effects in close proximity to neighbors. Our results show a lower bar fraction in less populated environments with larger distances to the first neighbor, suggesting that strong tidal interactions in LSBs could favor the bar presence. In this regard, observational studies by \cite{thompson1981bar} and \cite{andersen1996distribution}, analyzing the Coma and Virgo clusters, respectively, concluded that the bar fraction increases in regions with strong tidal interactions. Simulations by \cite{lokas2016tidally}, \cite{pettitt2018bars}, and \cite{smith2021brought} reinforce the idea that tidal interactions induce stronger, longer bars, forming earlier (\citealt{lokas2016tidally}), a mechanism that could be in place in the case of LSBs, especially in the case of close, massive perturbers that are able to affect the central region of the host galaxy (\citealt{kumar2021galaxy}). \\

\subsection{Bar fraction for control samples.}

The parameter distributions in HSB and LSB galaxy subsamples (Fig. \ref{pardist}) reveal that, in comparison with HSBs, LSBs are characterised by higher gas content, spin parameters, and tidal forces, but smaller $B/T$ values. Analyzing physical parameters for a fixed stellar mass (Figs. \ref{subfgasMs}, \ref{subflambdaMs}, and \ref{subBTMs}) shows that, especially in the low stellar mass range, LSBs exhibit higher gas content and spin values than HSBs. As stellar mass increases, the differences between the subsamples diminish, with HSBs approaching but not surpassing the gas fraction and spin parameters of LSBs. Remarkably, the $B/T$ ratio in LSBs remains consistently lower than in HSBs. Given that high gas content and spin negatively impact the bar fraction (similar to the bulge-to-total parameter), the observed low bar fraction in LSBs can be attributed to these characteristics. The distinct parameter distributions and their behavior at fixed stellar mass explain the differences in bar fractions between LSBs and HSBs, emphasizing the importance of scrutinizing the effects of these parameters on the bar fraction across all mass ranges, which we explore by making use of carefully constructed control samples. \\

Building upon the earlier discussion, we developed seven control samples (CS), all rooted in the LSB subsample. For each LSB galaxy, we selected a set of HSBs with assigned statistical weights, ensuring that the parameter distributions between the two subsamples are statistically similar. The primary aim of the control samples is to explore whether minimizing parameter differences can eliminate the observed disparity in bar fractions between LSB and HSB subsamples, as depicted in Fig. \ref{fbar-logM}, particularly in the low and intermediate stellar mass range. \\

The initial six control samples (CS1 to CS6) exclusively considered physical parameters, including stellar mass, gas mass fraction, bulge-to-total mass ratio, and spin parameter. In contrast, CS7 incorporated the external tidal force parameter $Q$. The controlled parameters for each CS are outlined in Table \ref{tab1}, and the results from CS1 to CS6 are presented in Fig. \ref{CS}. \\

In the case of the control sample CS1 (Fig. \ref{fig:a}), where only stellar mass is controlled, HSBs exhibit a higher bar fraction than LSBs. Beyond stellar masses higher than $10^{11}$ M${\odot}$, the bar fraction does not exhibit a significant change compared to the uncontrolled sample. When we control for other physical parameters in CS2, CS3, and CS4 (Fig. \ref{fig:b}, \ref{fig:c}, \ref{fig:d}, respectively), we observe a decrease and, in some cases, disappearance of the difference in bar fraction for the low and intermediate-mass range, especially in Fig. \ref{fig:d} (corresponding to CS4) where we simultaneously control for gas mass fraction and the spin parameter. In all three cases, a difference in bar fraction emerges (with LSBs presenting higher values) not observed in the uncontrolled sample in the high mass range. This suggests that when both parameters are controlled, LSBs and HSBs exhibit the same bar fraction for M$_{\star} < 10^{11}$ M$_{\odot}$, but for higher masses, these parameters introduce an artificial discrepancy between the subsamples. This combined impact of these two parameters was also studied by \cite{beane2023stellar}, who, using an N-body simulation of a Milky Way-like galactic disc, found that while the dark matter halo can slow down the bar due to angular momentum exchange, this effect is hindered by the presence of gas in the galaxy, which provokes that pattern speed to remain stable and constant. The gas can also accelerate the bar by providing a source of positive torque. They also found that the bars should still be able to decelerate below a certain gas fraction. \\

The most significant convergence in the bar fractions in the high stellar mass range above $10^{11}$ M$_{\odot}$ is observed in CS5 and CS6 (\ref{fig:e}, \ref{fig:f}) where the bulge-to-total fraction is controlled. In CS5, where only the bulge-to-total ratio ($B/T$) is controlled, we observe that for low and intermediate-mass ranges exists a difference between subsamples LSBs and HSBs. However, this difference disappears as we move towards larger masses when subsamples have the same bar fraction value, reaching $f_{\rm bar} \sim 0.6$. In the case of CS6, where both $B/T$ and gas mass fraction are controlled, we notice that the difference in bar fractions in the low stellar mass decreases compared to the uncontrolled sample. From the range of intermediate masses, we present a collapse in the bar fractions of LSBs and HSBs. We also do not see a difference at high masses. This suggests that for the intermediate and high mass ranges the bulge fraction and the gas play a significant role in the bar fraction and can make the discrepancy disappear in these ranges.

Finally, our last control sample CS7 exclusively considers the tidal parameter $Q$ (Fig. \ref{CS-Q}). The result of this study shows that within the low and intermediate-mass ranges, HSBs exhibit systematically higher bar fractions than LSBs. However, for the high mass range ($\geq 10^{11}$ M$_{\odot}$), an inverse trend appears, with LSBs showing higher bar fraction values than their HSB counterparts. These findings highlight a substantial difference between our subsamples for all mass ranges, suggesting that the tidal parameter $Q$ does not appear to have a significant effect on the presence of stellar bars. Even at large masses, it leads to a discrepancy not present in the original sample. 

It is important to emphasize that while the agreement between our simulation results and observational findings provides valuable insight, both simulations and observations are subject to various biases and limitations. The power of simulations lies not in their ability to serve as definitive proof, but in their capacity to trace the underlying processes that drive the observed phenomena—processes that remain largely inaccessible through direct observations.
\\

\subsection{Future prospects}

The TNG100 simulation has enabled us to investigate the influence of the galaxy physical and external parameters on the bar fraction for LSBs and HSBs, as previously examined in the section \ref{results}. 
The TNG100 simulation can resolve the physics above scales of approximately $1$ kpc due to its resolution. Although this is sufficient for a general examination of the formation and evolution of bars within the most massive galaxies in the local universe, simulations with a higher resolution are required to correctly track the initial phases of bar build-up at high redshifts. Furthermore, our study is based on a single snapshot of the simulation at $z=0$, and as such, it does not focus on the detailed formation and subsequent evolution of bars over cosmic time. To better understand the full evolutionary path of bars, simulations that include multiple snapshots over time and higher resolution would be necessary to capture the dynamic processes involved in bar formation and evolution.
For example, \cite{bi2022emergence,  bi2022modeling}, have investigated this issue through high-resolution zoom-in simulations of galaxies that are located in highly concentrated environments at high redshifts. A cosmological context would be an intriguing setting in which to conduct this experiment with a more extensive statistical galaxy sample.

It is crucial to investigate the fate of the gas in the nuclear region, which is influenced by the presence of a bar and is distinct from other physical processes such as SN and AGN feedback. This aspect is not addressed in this work, but it could be interesting to observe the formation of gaseous bars (\citealt{englmaier2004dynamical}) as a result of the torques generated by the stellar bar (\citealt{spinoso2017bar}).

It is also important to investigate the formation and evolution of bars in a wide range of galaxy stellar masses and in various environments. Tidal interactions and mergers may significantly influence the formation and weakening of bars. For example, the research conducted by \cite{lokas2016tidally} and \cite{lokas2014adventures} investigates the process of bar evolution in dwarf galaxies and Milky Way galaxies that arise from tidal interactions within a cluster-like environment. The authors found that the characteristics of the bars are subject to temporal variations and are influenced by the magnitude of the tidal force encountered during the evolutionary process. It is our intention to investigate this phenomenon in a future project using these cosmological simulations along with the TNG50 simulation, which is a higher-resolution simulation part of the TNG project (\citealt{pillepich2019first, nelson2019first}).

Lastly, the significant impact of local feedback processes, such as those from SNe or AGN, on the physical conditions of galaxies seems to establish favorable or unfavorable conditions for bar formation (\citealt{zana2019barred, bi2022emergence, bi2022modeling, rosas2024galaxy}), especially for different types of galaxies such as LSBs and HSBs. The parallel analysis of large cosmological simulations with varying feedback prescriptions could provide essential insight into the manner in which local processes interact with the conditions established by the large-scale environment and affect the formation and evolution of bars.

Although this study does not aim to explore the process of bar formation itself, it aims to understand how various physical parameters and the local environment correlate with the presence of bars in LSBs and HSBs.  This work provides insights into how the physical parameters of a host galaxy influence the presence of stellar bars in LSBs and HSBs, leveraging a self-consistent cosmological simulation to uncover potential physical mechanisms behind these observational findings.

\section{Conclusions}
\label{conclusions}

In this work, we studied the presence of stellar bars in low and high surface brightness (LSB and HSB) galaxies using the TNG100 simulation from the IllustrisTNG project to construct a sample consisting of  $4,244$ disc galaxies at $z=0$  with a stellar mass  M${_\star} \geq 10^{10}$ M$_{\odot}$. We found a lower bar fraction in LSBs ($24 \pm 1.73 \%$) when compared to HSBs ($28 \pm 0.74 \%$), particularly at low and intermediate stellar masses. This trend persists even when controlling for several key physical parameters and environmental factors, although the strength and direction of the difference varies depending on the parameter considered.

We showed that stellar mass alone does not fully account for the difference in bar fraction between LSBs and HSBs. Control samples matched in gas mass fraction and spin parameter reveal that these physical properties have a significant influence on bar presence, largely explaining the difference at low and intermediate masses. Interestingly, when both gas content and spin are jointly controlled (Fig. \ref{fig:d}), the bar fractions of LSBs and HSBs converge for $M_\star < 10^{11}$ M$_\odot$, suggesting that these two parameters are the main drivers of the lower bar incidence in LSBs in this regime.

At higher stellar masses, however, the discrepancy in bar fraction between LSBs and HSBs re-emerges, pointing to additional influences. We found that the bulge mass fraction becomes important at the high-mass end, where a more prominent bulge correlates with a higher bar fraction. This is consistent with results from the control sample that includes bulge prominence, which shows reduced differences in bar fraction between LSBs and HSBs at high masses (Fig. \ref{fig:e}).

Furthermore, we investigated the role of the environment using both the distance to the nearest neighbor and a tidal interaction parameter. LSBs show a stronger response to environmental effects, with their bar fraction increasing significantly in denser environments and under stronger tidal interactions. In contrast, HSBs exhibit a weaker or flat dependence on environment (Fig. \ref{environmentplots}). Our final control sample (CS7), which controls for the tidal parameter, confirms that environment contributes to the difference in bar fraction, though it does not fully account for it, especially at intermediate and high stellar masses (Fig. \ref{CS-Q}).

Overall, our results suggest that the lower bar fraction in LSBs arises from a combination of intrinsic physical properties—primarily higher gas fractions and spin parameters—and, to a lesser extent, environmental factors. At high stellar masses, bulge prominence becomes a key factor. These findings highlight the complex interplay between internal dynamics and external influences in shaping bar formation, and underscore the importance of carefully constructed control samples to isolate the effects of individual parameters.


\section*{Acknowledgements}

We thank the anonymous referee for a thorough reading of the original manuscript and for the insightful report that helped improve the quality of the article and the clarity of the results presented. Karol Chim-Ramirez and Bernardo Cervantes Sodi acknowledge the financial support provided by PAPIIT projects IN108323 and IN111825 from DGAPA-UNAM. Karol Chim-Ramirez also acknowledges the support of the SECIHTI scholarship and thanks Daniel J. Díaz González for his valuable technical support. Yetli Rosas Guevara acknowledges the support of the ``Juan de la Cierva Incorporation'' Fellowship (IJC2019-041131-I). Silvia Bonoli acknowledges support from the Spanish Ministerio de Ciencia e Innovación through project PID2021-124243NB-C21

\section*{Data Availability}

The data using in this work come from the IllustrisTNG simulations that are available in https://www.tng-project.org (\citealt{nelson2019illustristng}).



\bibliographystyle{mnras}
\bibliography{mnras_template}

\section{APPENDIX}
\label{anexes}
\begin{figure*}
\captionsetup[subfigure]{labelformat=empty}
\centering
\begin{subfigure}{0.3\linewidth}
    \includegraphics[width=\linewidth, height=0.8\linewidth]{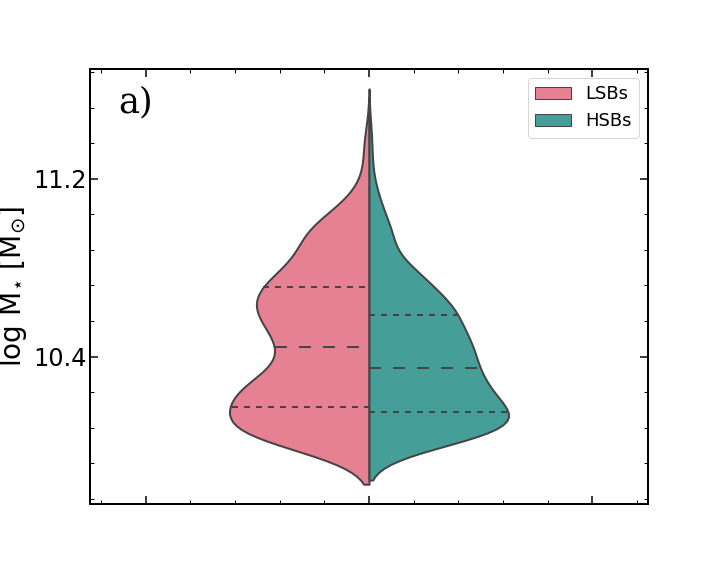}
    \caption{ }
    \label{Ms_ori}
\end{subfigure}
\hspace{0.5cm}
\begin{subfigure}{0.3\linewidth}
    \includegraphics[width=\linewidth, height=0.8\linewidth]{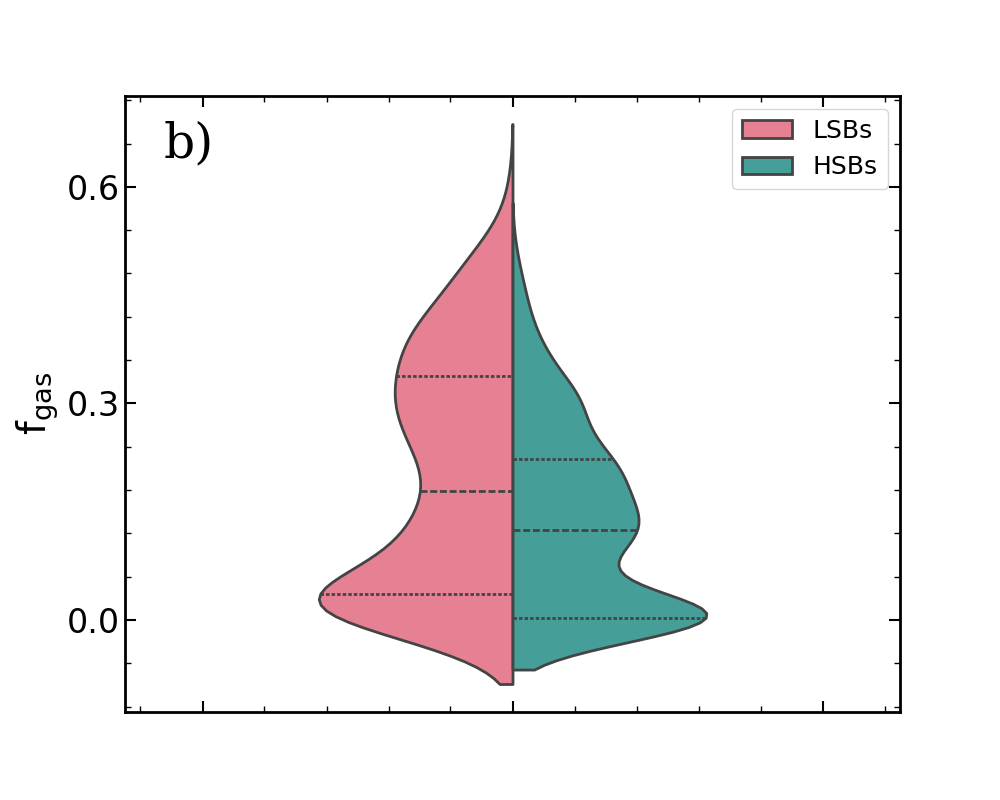}
    \caption{ }
    \label{Mgas_ori}
\end{subfigure}
\hspace{0.5cm}
\begin{subfigure}{0.3\linewidth}
    \includegraphics[width=\linewidth, height=0.8\linewidth]{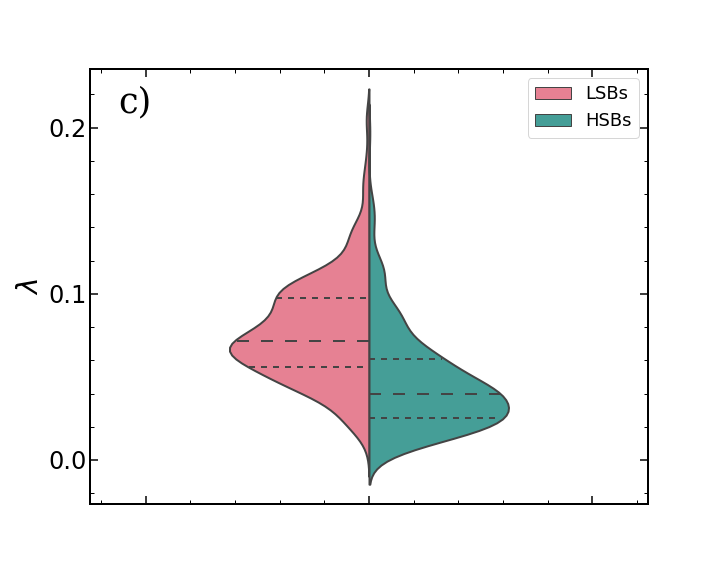}
    \caption{ }
    \label{lambda_ori}
\end{subfigure}

\vspace{0.5cm}

\begin{subfigure}{0.3\linewidth}
    \includegraphics[width=\linewidth, height=0.8\linewidth]{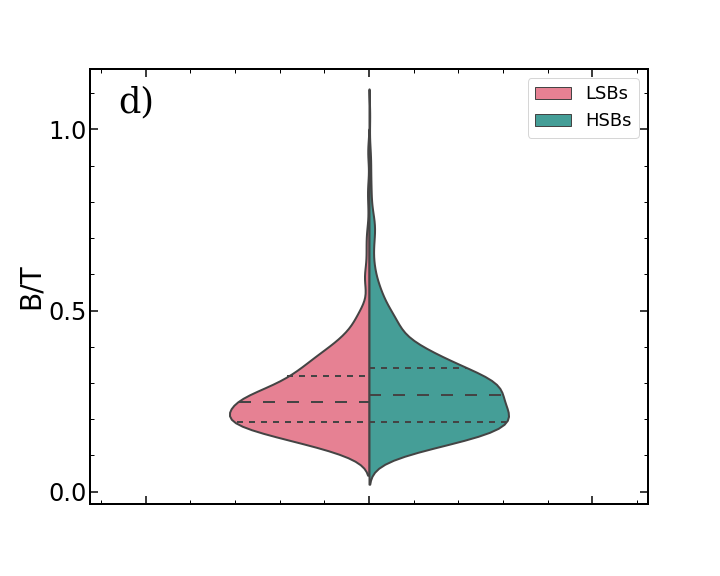}
    \caption{}
    \label{bulge_ori}
\end{subfigure}
\hspace{0.5cm}
\begin{subfigure}{0.3\linewidth}
    \includegraphics[width=\linewidth, height=0.8\linewidth]{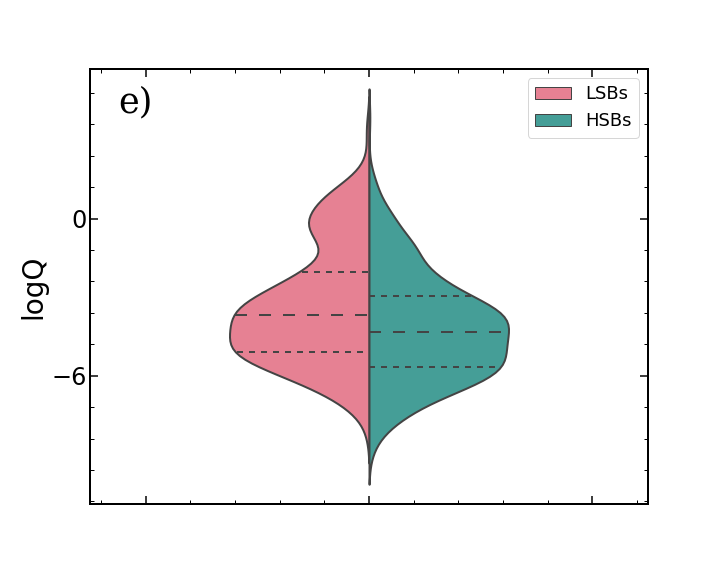}
    \caption{}
    \label{Q_ori}
\end{subfigure}

\caption{Probability distribution functions (PDFs) of stellar mass (\ref{Ms_ori}), gas mass fraction (\ref{Mgas_ori}), spin parameter (\ref{lambda_ori}), bulge-to-total mass fraction (\ref{bulge_ori}), and tidal parameter (\ref{Q_ori}) for the uncontrolled samples of LSBs and HSBs. For the spin parameter distribution, only central galaxies are considered. Dashed lines represent the quartiles of distributions.}
\label{IP_uncontrolled}
\end{figure*}

\begin{figure*}
\captionsetup[subfigure]{labelformat=empty}
\centering
\begin{subfigure}{0.3\linewidth}
    \includegraphics[width=\linewidth, height=0.9\linewidth]{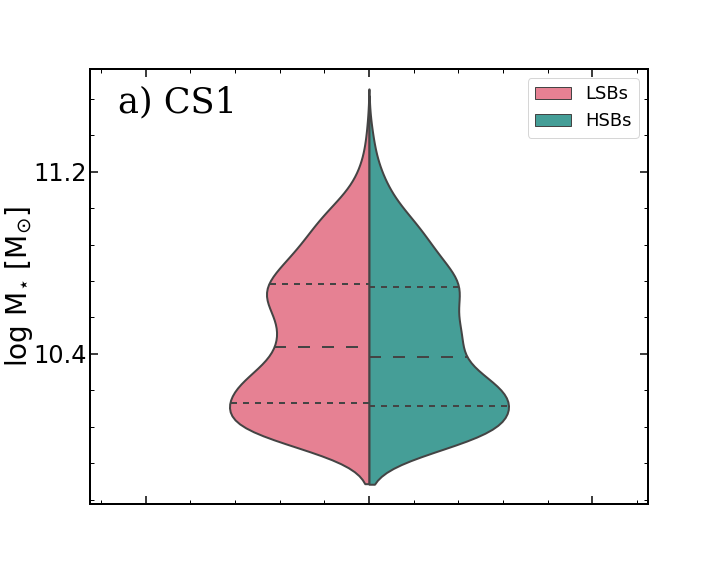}
    \vspace{-0.6cm}
    \caption{ }
    \label{CS1Ms}
\end{subfigure}
\hspace{0.01\linewidth}
\begin{subfigure}{0.3\linewidth}
    \includegraphics[width=\linewidth, height=0.9\linewidth]{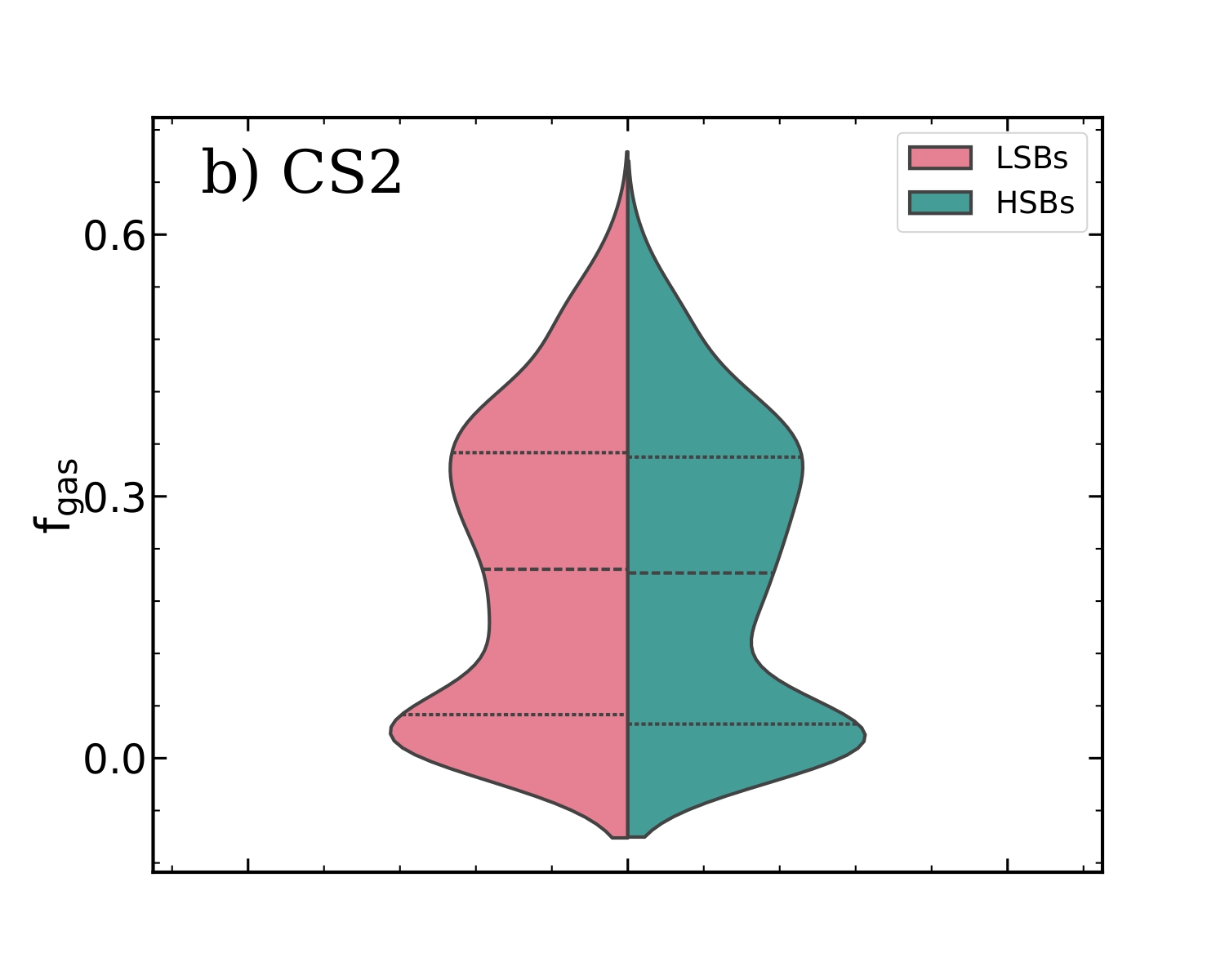} 
    \vspace{-0.6cm}
    \caption{ }
    \label{CS2Mgas}
\end{subfigure}
\hspace{0.01\linewidth}
\begin{subfigure}{0.3\linewidth}
    \includegraphics[width=\linewidth, height=0.9\linewidth]{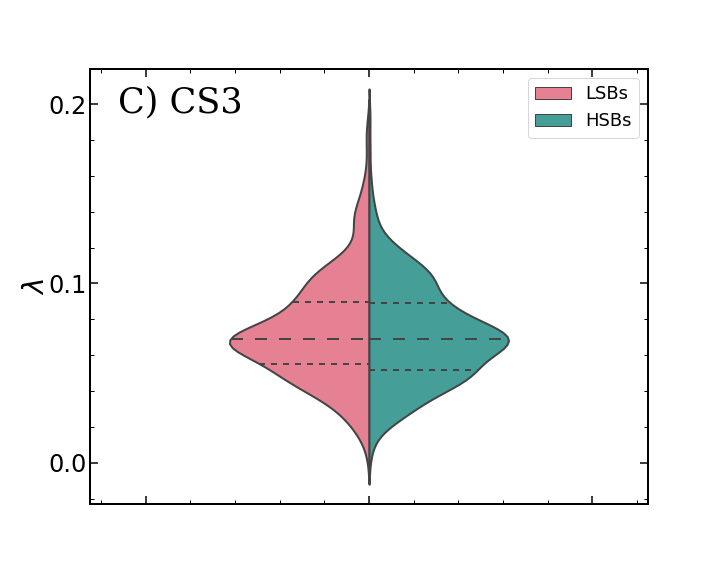}
    \vspace{-0.6cm}
    \caption{ }
    \label{CS3lambda}
\end{subfigure}

\hspace{-0.45cm}
\vspace{0.4cm}
\begin{subfigure}{0.36\linewidth}
    \includegraphics[width=1.6\linewidth, height=0.75\linewidth]{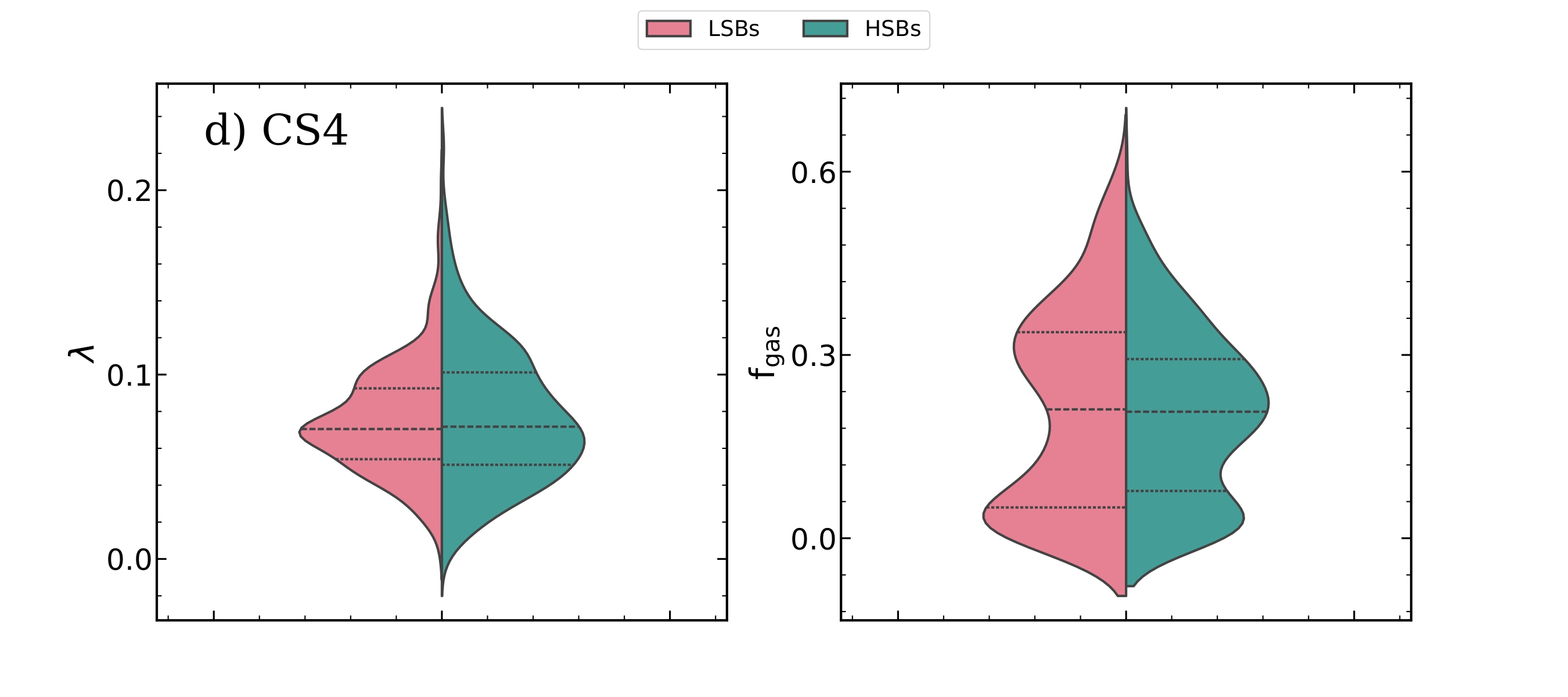} 
    \vspace{-0.3cm}
    \caption{ }
    \label{CS4lambdaMgas}
\end{subfigure}
\hspace{3.5cm}
\begin{subfigure}{0.3\linewidth}
    \includegraphics[width=\linewidth, height=0.9\linewidth]{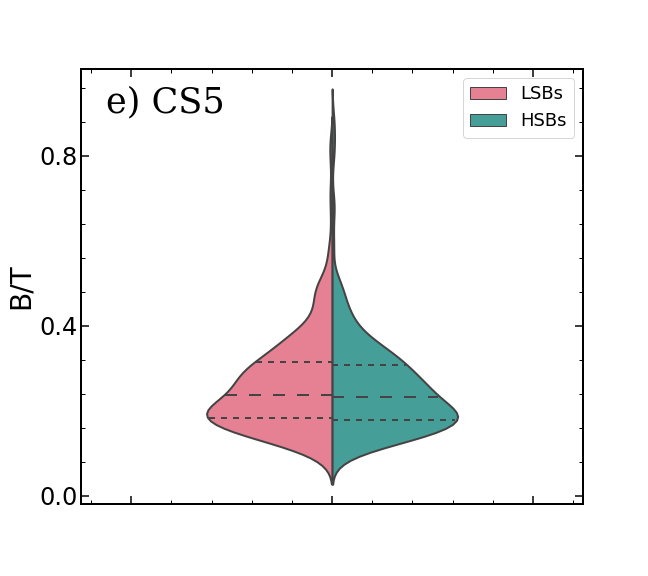}
    \vspace{-0.3cm}
    \caption{ }
    \label{CS5bulge}
\end{subfigure}

\vspace{-0.2cm}
\begin{subfigure}{0.36\linewidth}
    \includegraphics[width=1.6\linewidth, height=0.75\linewidth]{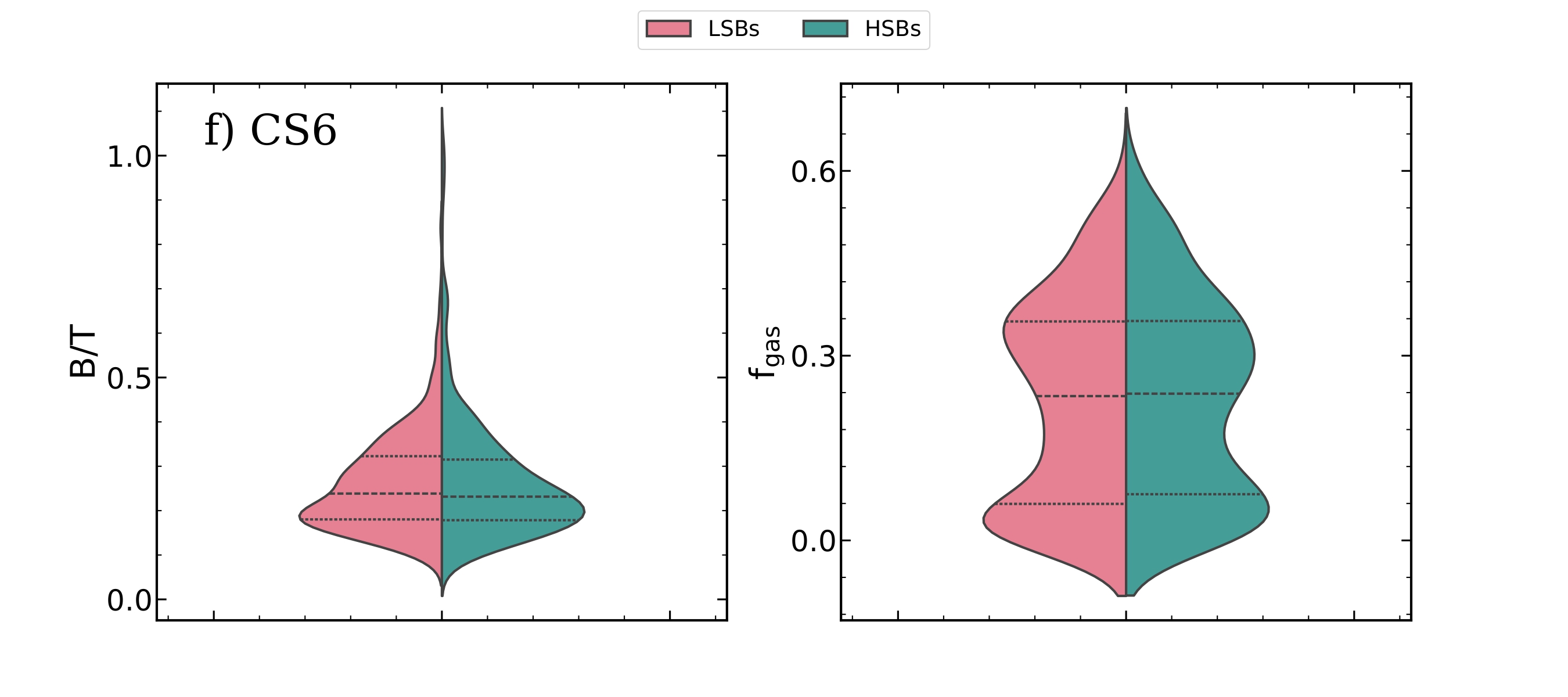} 
    \vspace{-0.3cm}
    \caption{ }
    \label{CS6bulgeMgas}
\end{subfigure}
\hspace{3.5cm}
\begin{subfigure}{0.3\linewidth}
    \includegraphics[width=\linewidth, height=0.9\linewidth]{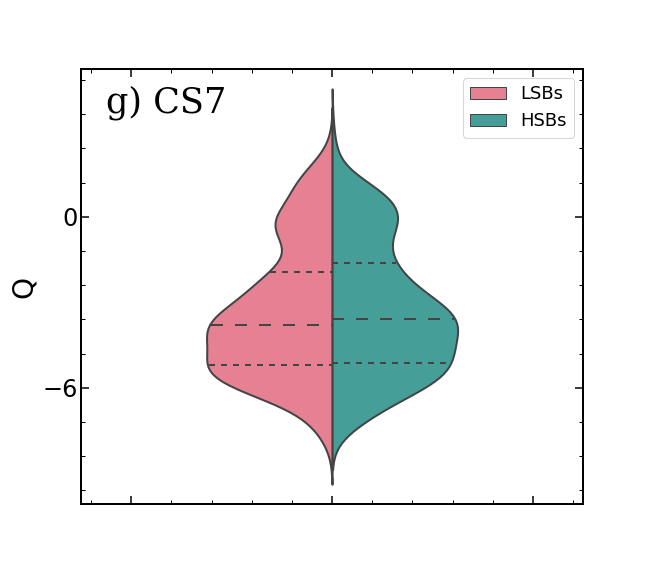}
    \vspace{-0.3cm}
    \caption{ }
    \label{CS7Q}
\end{subfigure}

\caption{Weighted probability distribution functions for the different control samples. Panels \ref{CS1Ms}, \ref{CS2Mgas}, \ref{CS3lambda}, \ref{CS5bulge} correspond to control samples where we controlled only by stellar mass, gas mass fraction, spin parameter or bulge, respectively. Panels \ref{CS4lambdaMgas} and \ref{CS6bulgeMgas} present the distribution for the control samples that consider a couple of parameters: spin and gas mass fraction; and bulge and gas mass fraction, respectively. Finally, in \ref{CS7Q} we control by tidal parameter. Dashed lines correspond to the quartiles of distributions.}
\label{PDF_control_samples}
\end{figure*}

In this section, we include the probability distribution function (PDF) of the physical parameters analyzed in this work. Figure \ref{IP_uncontrolled} presents the parameter distributions for the uncontrolled LSB and HSB samples, dashed lines correspond to the quartiles of distributions. As we can see, the original distributions are different for LSBs and HSBs. On the other hand, figure \ref{PDF_control_samples} presents the corresponding weighted distributions for our different control samples. As we can see, the corresponding parameter distributions become similar for the control samples, indicating that the control method works correctly. In addition, in Table \ref{tab:ks_values} we present the p-values obtained from  Kolmogorov–Smirnov (KS) tests performed when comparing each parameter for the control samples.

\begin{table}
    \centering
    \begin{tabular}{|c|c|c|}
        \hline
         & \textbf{physical parameter} & \textbf{p-value from KS test} \\
        \hline
        \textbf{CS1} & M$_{\star}$ & $0.84$ \\
        \hline
        \textbf{CS2} & f$_{gas}$ & $0.93$ \\
        \hline
        \textbf{CS3} & $\lambda$ & $0.97$ \\
        \hline
        \multirow{2}{*}{\textbf{CS4}} & f$_{gas}$ & $0.86$ \\
         & $\lambda$ & $0.95$ \\
        \hline
        \textbf{CS5} & $B/T$ & $0.94$ \\
        \hline
        \multirow{2}{*}{\textbf{CS6}} & f$_{gas}$ & $0.89$ \\
         & $B/T$ & $0.94$ \\
        \hline
        \textbf{CS7} & $Q$ & $0.91$ \\
        \hline
    \end{tabular}
    \caption{Table with p-values for physical parameters for each control sample.}
    \label{tab:ks_values}
\end{table}

\bsp	
\label{lastpage}
\end{document}